\documentclass[a4paper]{article}
\usepackage{siunitx}

\usepackage{cite}
\usepackage{amsfonts}

   \usepackage[pdftex]{graphicx}
\usepackage{amsmath}
\usepackage{algorithm}
\usepackage{algpseudocode}

\usepackage{array}
\usepackage{multirow}

\usepackage{subfigure}
\newcommand{\argmin}{\operatornamewithlimits{argmin}}

\begin{document}
%
\title{Learned SIRT for Cone Beam Computed Tomography Reconstruction}
%
%
%

\author{Roeland J. Dilz, Lukas Schr\"oder, Nikita Moriakov, \\ Jan-Jakob Sonke and Jonas Teuwen
\thanks{R. J. Dilz, L. Schr\"oder, J.-J. Sonke and J. Teuwen are with the Netherlands Cancer Institute, Amsterdam, The Netherlands, Plesmanlaan 121, 1066 CX Amsterdam. email:j.sonke@nki.nl. N. Moriakov is with Radboud University, Nijmegen, The Netherlands, PO Box 9102 6500HC Nijmegen}}

%
%

\markboth{Preprint for IEEE Transactions on Medical Imaging}%
{Shell \MakeLowercase{\textit{et al.}}: Bare Demo of IEEEtran.cls for IEEE Journals}
%



\maketitle

\begin{abstract}
We introduce the learned simultaneous iterative reconstruction technique (SIRT) for tomographic reconstruction. The learned SIRT algorithm is a deep learning based reconstruction method combining model knowledge with a learned component. The algorithm is trained by mapping raw measured data to the reconstruction results over several iterations. The Learned SIRT algorithm is applied to a cone beam geometry on a circular orbit, a challenging problem for learned methods due to its 3D geometry and its inherent inability to completely capture the patient anatomy. A comparison of 2D reconstructions is shown, where the learned SIRT approach produces reconstructions with superior peak signal to noise ratio (PSNR) and structural similarity (SSIM), compared to FBP, SIRT and U-net post-processing and similar PSNR and SSIM compared to the learned primal dual algorithm. Similar results are shown for cone beam geometry reconstructions of a 3D Shepp Logan phantom, where we obtain between 9.9 and 28.1 dB improvement over FBP with a substantial improvement in SSIM. Finally we show that our algorithm scales to clinically relevant problems, and performs well when applied to measurements of a physical phantom.
\end{abstract}


%

\section{Introduction}
Computed tomography (CT) is a commonly used imaging technique in medicine where a series of X-ray measurements from different angles are acquired with the goal of finding the distribution of attenuation coefficients of the underlying tissue. 
These images, or projections, can be obtained with different acquisition strategies. In th is paper, we focus on a cone beam (CB) geometry with a single circular orbit, where each of these images is acquired with a flat panel detector and are in essence 2D projections at a certain angle. Cone beam CT (CBCT) plays an important role in many fields of medicine, including dentistry, interventional radiology, surgery and radiation oncology. While our work is applicable to all such applications of CBCT, we are mainly inspired by its applications to radiation oncology where CBCT is used as a means of treatment verification and adaptation.

In radiation oncology, or radiotherapy for short, radiation is delivered in one or several fractions over the period of several weeks to target malignant tissue. To verify the patient position with respect to the treatment plan a daily CBCT scan is acquired\cite{Jaffray02,Sonke19}. Ideally such a scan could be used to \emph{adapt} the treatment plan to the current patient anatomy and tumor response. However, this is limited in the current CBCT acquisition by the poor soft-tissue contrast, and non-calibrated intensity values which are required to compute the accumulated dose to the tumor and healthy tissue. There are several underlying factors for this. First of all, the cone beam geometry is inherently unable to capture the patient anatomy completely due to the Defrise or cone beam artifact \cite{defriseartifact}. 
Secondly, due to the large panel size in CBCT, scatter \cite{Zhu09}, ghosting and image lag \cite{Mail07} play an important role. Finally, the slow acquisition time, typically in the order of minutes, can cause misalignments in the projections due to anatomical motion such as the respiratory motion \cite{Sonke05}. While this can be approached by binning the projections in respiratory bins, this results in a significantly non-uniformly subsampled problem for each phase in the respiratory cycle. Our algorithm is designed to alleviate the Defrise artifact and reconstruct images from a low number of projections.

The aim of this paper is to present a deep learning based reconstruction method for CBCT which achieves good soft-tissue contrast, e.g., to be able to distinguish tumor from healthy tissue before the delivery of radiation.

Reconstruction methods, where anatomy is inferred, belong to the class of inverse problems which are subject of active research. In terms of deep learning based methods three parallel approaches can be distinguished: (i) learned post-processing where a classical (i.e., non-deep learning) reconstruction method is post-processed by a neural network trained to remove artifacts from the reconstructed image; (ii) a full data-driven approach where a neural network maps the raw input data directly to the reconstruction and finally (iii) learned schemes which combine information about the forward operation (and therefore part of the physics) with a neural network. Our method fits in the latter category. In contrast to (i) such an approach uses all the information available in the measurements, whereas a learned post-processing merely attempts to filter the artifacts created by the reconstruction method. Our approach, the learned SIRT (lSIRT) algorithm, in contrast to (ii) combines knowledge of the underlying physics with a neural network to improve the final reconstruction. A complete data-driven approach would need to learn this forward operator, and to our knowledge so-far no such method for computed tomography has been demonstrated which scales to clinically relevant problems. 

\subsection*{Contribution and overview of paper}
In this paper, we introduce the Learned SIRT (lSIRT) algorithm for CBCT. lSIRT provides superior soft-tissue contrast and has less artifacts compared to the algorithms for CBCT reconstructions in current clinical use. Due to its design, the lSIRT algorithm is applicable to both 2D and 3D problems, and readily scales to clinically relevant sizes while requiring modest computational resources. 
Our method only uses neural networks in the image domain. This has the advantage that the network can be trained and tested on large input sizes by tiling the input, i.e. by dividing the input in a set of smaller regions that can be handled subsequently. This is not possible using deep learning based reconstruction methods which learn both in the projection as in the image domain \cite{Adler18}, limiting their extension from the 2D setting to the 3D CBCT domain.

Next to showing that our method performs well in 3D CBCT reconstruction, we also show that in 2D our method leads to competitive results when compared to other learned reconstruction methods. Finally, we show the applicability of the lSIRT algorithm to  measurements of a physical phantom.

\section{The Learned SIRT algorithm}
\subsection{Inverse problems and regularization}
Image reconstruction problems can be formulated as an inverse problem. From a functional analytic viewpoint, an inverse problem is posed as follows: given an image $x \in X$ and measured data $y \in Y$ we write
\begin{equation}
y = A x + \eta
\end{equation}
where $A: X \to Y$ is the forward operator, or projection operator, that models how the data $x$ gives rise to measurement $Ax$ in the absence of noise, and $\eta$ is an $Y$-valued random variable modeling the noise component of the measurements. The measurements in $Y$ are often referred to as \emph{projections}, or, in the case of CT reconstruction, as \emph{sinograms}. Typically, the spaces $X$ and $Y$ are Banach or Hilbert spaces, and in our case these are spaces of functions describing true anatomy and measurements. Compare this to a Bayesian perspective, where $X$ and $Y$ are probability spaces and the probability distribution $x \sim P(x)$ on $X$ is called the \emph{prior}. Bayes theorem states that
\begin{equation}
P(x|y) = \frac {P(y|x)P(x)} {P(y)}, 
\end{equation}
where the conditional probability $P(y|x)$ is called the \emph{likelihood}, which expresses the probability of measurement $y$ with data $x$ given, and is derived from the forward model.

The goal of reconstruction is to retrieve the image $x$ from the noise-corrupted measurements $y$. Inversion of the operator $A$ is generally an ill-posed problem. There are several reasons for this. If the linear operator $A$ has a nontrivial kernel $\ker A$, then its inverse is not uniquely defined. Secondly, for infinite dimensional spaces the inverse can be unbounded, implying that small variations in measurement noise $\eta$ can lead to very different solutions. When working with finite-dimensional discretizations of the operator $A$, this finite-dimensional discretization can be poorly conditioned, which in practice can result in numerical instabilities. 

Some form of regularization is typically utilized in order to combat the ill-posedness of inverse problems. The goal of functional analytic regularization is, formally, to provide a parametrized mapping $R_{\theta}: Y \to X$ (existence of solutions) that
is continuous in $Y$ for fixed parameter $\theta$ (stability of solutions) and convergent in the sense that there is a way to select a sequence $( \theta_i )$ so that $R_{\theta_i} y_i \to x$ as $y_i \to Ax$. A particular approach to functional analytic regularization methods is given by the family of \emph{variational methods}. In variational methods, the regularization scheme is defined as
\begin{equation}
R_{\theta}(y) := \argmin_{x' \in X} \lbrace \mathcal L(A x', y) + S_{\theta}(x') \rbrace.
\label{eq.basicrec}
\end{equation}
The first term here is referred to as the \emph{data fidelity term}, the second as the \emph{regularization term} and $\theta$ is the parameter vector of the regularization term. A particular example would be \emph{Tikhonov regularization}, which is defined as
\begin{equation*}
R_{\theta}(y) := \argmin_{x' \in X} \lbrace \| A x' - y \|_2 + \theta \| x' \|_2 \rbrace,
\end{equation*}
where $\theta \geq 0$ determines the weight of the regularization term. Another important example for image reconstruction is the TV regularization, where the regularization term $S_{\theta}(x)$ is defined using total variation of $x$, i.e.,
\begin{equation*}
S_{\theta}(x) := \theta \cdot \mathrm{TV}(x) = \theta \int\limits_{\Omega} |\nabla x|(z) dz,
\end{equation*}
where $\Omega$ is the volume in which reconstruction is performed.

Many learned approaches to CT and MRI reconstruction are post-processing only \cite{Jin2017,Zbontar18,Chen17}, and combine a classical reconstruction operator $A^\dagger$ such as FBP with a learned post-processing operator $P$ to get the output $x = P A^\dagger y$. Alternatively, one could combine iterative schemes with a learned model \cite{Adler17,Adler18,Lonning19}. A potential advantage of such an approach is that the data can be used more efficiently. Our algorithm is based on a direct minimization of \eqref{eq.basicrec} via gradient descent. However, instead of an analytically defined regularization term $S_{\theta}$, we only make use of the gradients $\nabla_x S_{\theta}(x)$, which are learned by a neural network with parameters $\theta$.

\subsection{Maximum likelihood: classical SIRT}
When no prior knowledge is available on the object which we want to measure, we can take $P(x)$ to be constant for all images.
In this case $P(x|y)$, by Bayes theorem, is proportional to the likelihood function $P(y|x)$ divided by $P(y)$. Since $P(y)$ does not depend on $x$, this term can be ignored when optimizing over $x$ and we recover the maximum likelihood estimate \cite{liang89}.
The noise is Poisson-distributed in the pre-log measurement, which for high photon counts corresponds to Gaussian noise as a good approximation of the post-log attenuation values.
For such Gaussian noise, the noise distribution is independent of $y$ and characterized by the probability density function $P_{\sigma}(\eta) = \frac1{\sigma (2\pi)^{-k/2}} \exp( -\frac1{2\sigma^2} \|\eta\|^2_{L^2})$ where $\sigma$ is a parameter related to the intensity of the noise and $k$ is the dimension of projection space. Therefore, the conditional probability $P(x|y)$ is proportional to $P(y|x)$ with proportionality constant that does not depend on $x$, and $P(y|x)$ equals $P_{\sigma}(A x - y)$. To summarize,
\begin{equation}
   P(x|y) \simeq \frac1{\sigma (2\pi)^{-\frac k 2}} \exp\left( -\frac1{2\sigma^2} \|A x - y\|^2_{L^2} \right).
\label{eq.sirtprob}
\end{equation}

Maximizing the quantity on the right-hand side is computing the maximal likelihood, i.e., we maximize the quantity with respect to the unknown image $x$, so that the measurement $y$ corresponds to the most probable signal. Taking the logarithm and minimizing with gradient descent with step size $\lambda / 2$ we find
\begin{equation}
\begin{split}
x^{(0)} &= 0,\\
x^{(k+1)} &= x^{(k)} + \lambda A^T(A x^{(k)} - y).
\end{split}
\label{eq.sirt1}
\end{equation}
This is the Simultaneous Iterative Reconstruction Technique (SIRT) algorithm \cite{Gilbert72,Dines1979,sluis90,Kak02}. 
For sufficiently small $\lambda$ the iterative scheme is convergent, and for a sufficiently large number of iterations gives a good approximation to $x$.

Although it is feasible to work with this SIRT variant, in another variant of SIRT the step size $\lambda$ does not have to be chosen explicitly. For this, we use \cite[Eq.~4.1 and 4.2]{sluis90} with $\alpha = \omega = 1$, such that
\begin{equation}
\begin{split}
x^{(k+1)} &= x^{(k)} + C  A^T R (A x^{(k)} - y)\\
C_{jj} &= 1/\sum_{j}^{}a_{ij} \quad \quad \quad 	R_{ii} = 1/\sum_{i}^{}a_{ij}	,
\end{split}
\label{eq.sirt}
\end{equation}
where $R_{ij} = C_{ij} = 0$ for $i \neq j$ and $a_{ij}$ are the individual components of $A$. These matrices $C$ and $R$ are diagonal matrices that contain the sum of the columns and rows of the projection matrix, respectively.

\subsection{Estimate based on posterior information}
In many cases some prior information is available.
For example, when we are taking a thorax CT, we can expect that the image resembles a CT scan of a thorax and not random noise.
This prior knowledge can be incorporated in the likelihood $P_X(x)$ of an image $x$, and \eqref{eq.sirtprob} is replaced with
\begin{equation}
\begin{split}
P(x|y) &\simeq P_X(x)P_\sigma(A x - y) \\
\simeq &\exp\Bigl( -\frac1{2\sigma^2} \|A x -y \|^2_{\mathcal L^2}+ \log P_X(x) \Bigr).
\label{eq.lsirtprob}
\end{split} 
\end{equation}
And the minimization procedure of \eqref{eq.sirt1} now becomes
\begin{equation}
\begin{split}
x^{(0)} &= 0 \\
x^{(k+1)} &= x^{(k)} + \alpha \nabla_x \log P_X(x^{(k)}) + \lambda A^T(A x^{(k)} - y),
\end{split}
\label{eq.sirt2}
\end{equation}
where $\alpha$ determines the rate of convergence towards the posterior.
The expression $\log P_X(x)$ can be viewed as a regularizer term in \eqref{eq.basicrec}, but an explicit analytic form of $\log P_x(x)$ is not available.

Instead we propose to employ a learned function $g_\theta(x)$ as a replacement for the gradients $\nabla_x \log P_X$ in \eqref{eq.sirt2}. In the derivation we relied on the following heuristics.
Suppose that $x^{(k)}$ is sufficiently close to $x_{\text{true}}$, which is, additionally, a local maximum of $P_{X}$. The first assumption holds after a sufficient number $k$ of classical SIRT iterations, because then the classical SIRT solution $x^{(k)}$ of  \eqref{eq.sirt1} is close to the true solution $x_{\text{true}}$. In this setting, we observe that $x_{\text{true}} - x^{(k)}$ should point approximately in the direction of the gradient $\nabla_x \log P_X(x)$.
Therefore, if we learn a function $g_\theta(x^{(k)}) \approx x_{\text{true}}$, this would allow to estimate the gradient
\begin{equation}
\nabla_x \log P_X(x) \approx  x_{\text{true}} - x^{(k)} \approx g_{\theta}(x^{(k)}) - x^{(k)}.
\label{eq.gintro}
\end{equation}
Now we can replace $x^{(k)} + \nabla_x \log P_X(x^{(k)})$ in \eqref{eq.sirt2} by $g_{\theta}(x^{(k)})$. 
For a small enough $\alpha$ this will lead to a stable algorithm.
We now find
\begin{equation}
\begin{split}
x^{(0)} &= 0 \\
x^{(k+1)} &= (1-\alpha) x^{(k)} + \alpha g_{\theta}(x^{(k)}) + \lambda  A^T(A x^{(k)} - y).
\end{split}
\label{eq.learnedsirtpre}
\end{equation}
Note that for $\alpha = 0$ we recover the classical SIRT algorithm of \eqref{eq.sirt1}. Furthermore, even if the original assumptions about $x_{\text{true}}$ being a local maximum does not hold, this update rule remains meaningful, since it essentially interpolates between SIRT update and the best neural network estimate of $x_{\text{true}}$.
Similarly as in \eqref{eq.sirt} we get
\begin{equation}
x^{(k+1)} = (1-\alpha) x^{(k)} + \alpha g_{\theta}(x^{(k)}; z)+ C  A^T R (A x^{(k)} - y).
\label{eq.learnedsirt}
\end{equation}
which is the procedure for the algorithm we will henceforth refer to as lSIRT. Note that a parameter $z$ was added in \eqref{eq.learnedsirt}. As $g_\theta$ is a neural network, this can be a convenient way to add other prior information such as previous iterates. In the learned primal dual (LPD) algorithm \cite{Adler18}, the iterative scheme is unrolled, and in effect the ``history" of previous iterations is propagated through the network. To do this, we add $x^{(k)}$ and $A^T (y-A x^{(k)})$ as extra channels, and train the neural network with a loss function \eqref{eq:loss-function} to get:
\begin{equation}
	g_\theta(x^{(k)},x^{(k-1)},A^T (y-A x^{(k)}))_i \approx \begin{cases} x_{\text{true}}, \\x^{(k+1)}- x_{\text{true}}. \end{cases}
	\label{eq.genh}
\end{equation}
where $\theta$ stands for a set of parameters that is trained such that \eqref{eq.genh} is approximated with sufficient accuracy. 
In Section~\ref{sec:lsirt-star} we will elaborate on the effect of these extra channels. 


\subsection{Learning for 3D CBCT}
Several deep learning based CT reconstructions have recently been proposed \cite{Adler17,Adler18,Chen17,Chen17-2}. While giving excellent results for 2D fan- and parallel beam geometries, the problem is far more pressing for cone beam geometries which are inherently 3D. Even more so, for 2D excellent reconstruction methods exist \cite{Lechuga16} and most 2D geometries, in contrast to the CB geometry, can sample the complete image domain. In CBCT this leads to the typical cone beam artifact  \cite[Section~5.1.2]{defriseartifact}. Furthermore, several of these state-of-the-art architectures achieve excellent results in 2D, but do not readily scale to clinically relevant problems in 3D as these either attempt to learn the forward projector, or perform learning both in the image and projection domain, increasing the memory requirement.

The lSIRT algorithm circumvents these two problems by combining domain knowledge in the form of the forward projector and by performing learning only in the image domain, thereby allowing for a patch based training. Such a patch based approach allows to trade off memory for computation. This allows to scale the problem to clinically relevant problems with only modest computational resources which are readily available in the clinic. To be clinically relevant, the volumes that can be reconstructed should be at least $256^3$ and the reconstruction should be fast and complete in e.g. a few minutes on standard hardware in the clinic. 

\section{Implementation and evaluation}
We evaluate the lSIRT algorithm both in 2D using a parallel beam geometry and in 3D using a cone beam geometry, both for simulated data and real phantom measurements.

In this section we describe the datasets used to train and evaluate the model, the model architecture and the implementation details.

\subsection{Data}
\subsubsection{Simulated data}
As ground truth we use four different datasets for which we simulate the projections. Depending on whether we build a model for 2D or 3D, we simulate projections for a parallel beam geometry in 2D, or for a 3D CBCT geometry with a source-to-axis distance of \SI{1}{\meter}, a source to detector distance of \SI{1.5}{\meter}, and a detector with a pixel pitch of \SI{1}{\milli\meter}. All reconstruction volumes have a \SI{1}{\milli\meter} pitch.

These are: (i) images with six randomly generated triangles in $128 \times 128$, the same reconstruction size as in \cite{Adler18}, where the per triangle intensity is randomly distributed according to a gamma distribution with scale $1$ followed by $L^2$ normalization; (ii) the $128 \times 128$ pixel Shepp-Logan phantom (validation only); (iii) images with 20 randomly generated ellipses in $128 \times 128 \times 128$ with the center uniformly sampled in the image, and radii sampled from the absolute value of a zero mean uniform distribution with variance $128 / 3$. Per ellipse intensity is sampled from a standard normal distribution; (iv) consists out of 42 (37 training, 5 testing) or 338 (308 training, 30 testing) CT scans for the 2D and 3D case respectively. The in-plane resolution for all scans was $512 \times 512$ with a variable number of slices and slice thickness ranging between \SI{1}{} and \SI{4}{\milli\meter}. The images were randomly selected from lung cancer patients treated at the radiation oncology department at our institute between 2015 and 2019. This study was approved by the local ethics board after summary review with waiver of full review and informed consent. The training set included thorax CTs and occasional head CTs, for patients where brain metastasis were treated. Since deep learning frameworks are often optimized for quantities with approximately unit amplitude, all CT scans were scaled by $10^{-3}$. We will henceforth refer to this data as the \emph{lung data}.

The number of detector elements and the number of angles depends on the data size and whether we work in 2D or 3D. The number of detector elements was chosen to fit the complete object on the detector. For 2D, we followed \cite{Adler18} and selected 185 detector elements and 30 angles for the triangles (i) and 742 detector elements and 120 angles for the lung data (iv). All projections are equidistantly sampled over 360 degrees. The numbers for the lung data are of similar proportion for the numbers for the triangles with respect to the image size.
Similar numbers were taken for the 3D CBCT geometry, where we reconstruct to $128^3$ and $256^3$, where $185^2$ detector elements with $30$ projections and $371^2$ and $60$ projections are selected, respectively.

\begin{figure}[t]
	\centering
	\subfigure[2D: triangles]                         {\includegraphics[width = 1.2in]{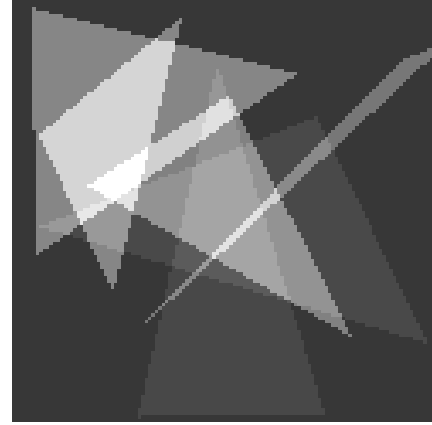}}
	\subfigure[3D CBCT: Ellipses]                         {\includegraphics[width = 1.2in]{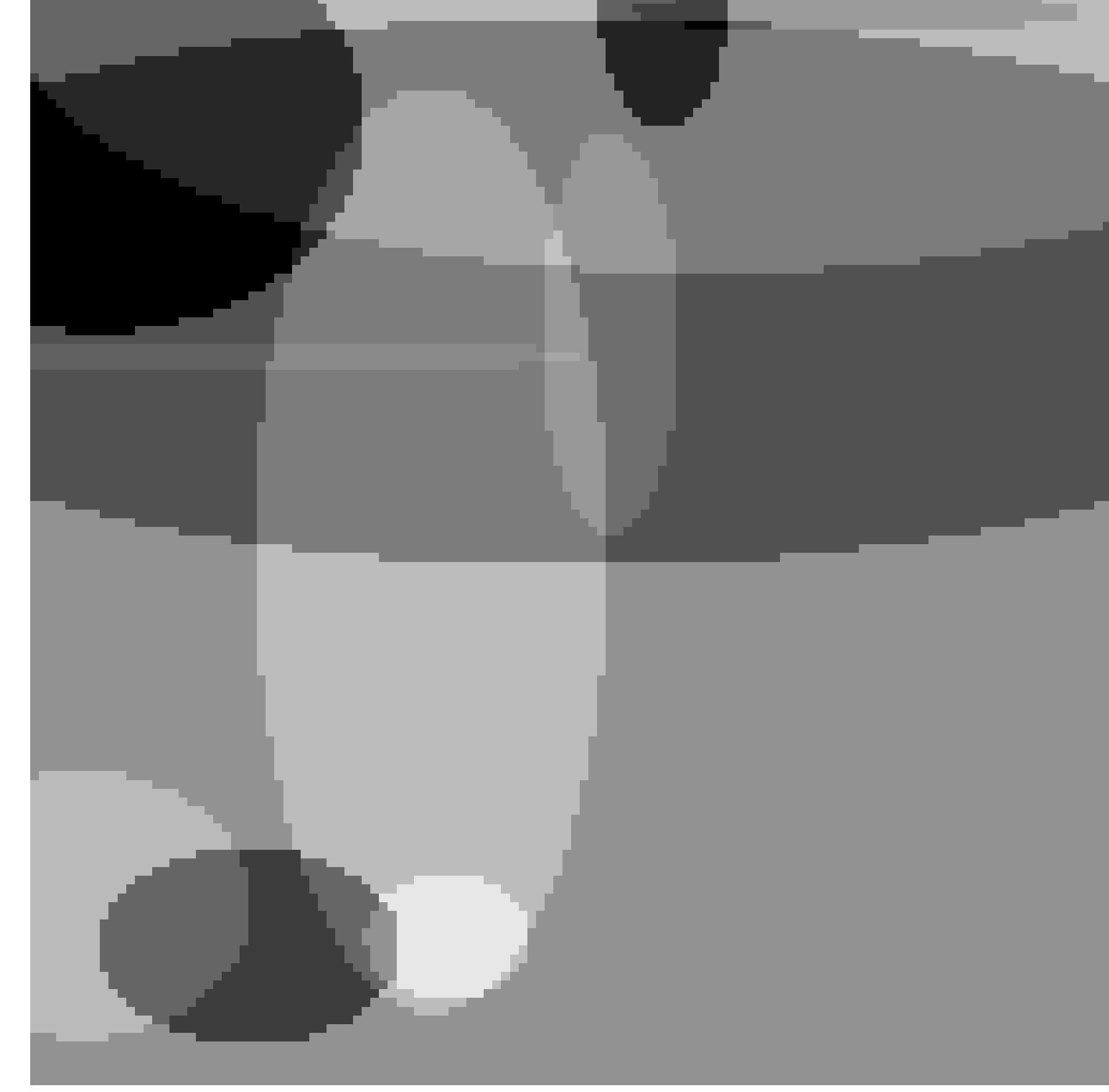}}
	\caption{Examples of artificial phantoms, with zero backgrounds.}
	\label{fig:phantom}
\end{figure}

Examples of the triangles (i) and ellipses (iii) are given in Figure~\ref{fig:phantom}.
For all data we added additive normally distributed noise to the projections with intensity $0.0025$, $0.0225$ and $0.0625$ to which we refer as the \emph{low noise}, \emph{medium noise} and \emph{high noise} regimes
respectively.

\subsubsection{3D phantom data}\label{sec:phantom-data}
To evaluate our model on measured data, a CIRS CBCT Electron Density and Image Quality phantom (CIRS Inc., Norfolk, Virginia, USA) was scanned on a linac integrated scanner (Synergy, XVI 5.0, Elekta Ltd, Crawley, UK) with the use of a bow-tie filter and an anti-scatter grid\cite{Uros14}. A flat-panel detector was utilized (XRD 1640 AL3 ES, PerkinElmer, Waltham, MA, USA). The field-of-view was \SI{25}{\centi\meter} and the phantom was assembled to resemble the head and neck region. The nominal values were a peak voltage of \SI{120}{\kilo\volt}, a tube current of \SI{16}{\milli\ampere}, and a pulse length of \SI{40}{\milli\second}. A full rotation with $342$ projections was measured, and their projection angles were recorded.

For the quality assessment we use a different metric for the measured data, as no ground truth is available.
The CT number inserts of the phantom, ranging from air to Teflon, which were positioned in the iso-center, were used for the image analysis. The polystyrene insert was chosen to evaluate the contrast-to-noise ratio (CNR) because its CNR is the lowest. A cylindrical volume-of-interest (VOI) in the insert and two VOIs next to the insert were chosen to calculate the CNR as
\begin{equation}
\text{CNR} = \frac{|\overline{\text{CT}}_i-\overline{\text{CT}}_s|}{\sqrt{\sigma_i^2+\sigma_s^2}}
\label{eq.CNR}
\end{equation}
where the subscripts $i$ and $s$ denote the insert and the surroundings respectively, $\overline{\text{CT}}$ are the mean CT numbers and $\sigma$ the standard deviations of the VOI. To analyze the spatial resolution the edge response of the Teflon insert in the central slice was used. After transforming a region of interest (ROI), which includes the Teflon, into polar coordinates and averaging along the angle, a fit in form of a cumulative normal distribution function was fitted to this edge response. The derivative of this edge response fit is the line spread function (LSF) and its full width at half maximum (FWHM) is a measure for the spatial resolution \cite[Chapter~25]{SmithDSP}.

\subsection{Network architecture and training}
\begin{figure}[ht]
	\centering
	\includegraphics[width=8.5cm]{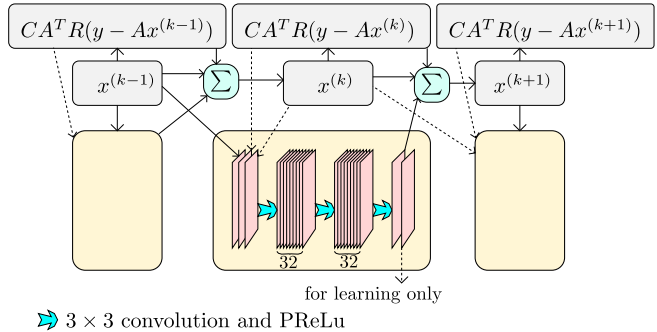}
	\caption{A graphic depiction of an iteration of the learned SIRT algorithm. All solid parts are present in our implementation of plain lSIRT*, described in \eqref{eq.learnedsirt}. The dotted parts are only present in the version of \eqref{eq.genh}.}
	\label{fig:lsirt}
\end{figure}

The neural network $g_\theta$ described by \eqref{eq.learnedsirt} was chosen to be a CNN and parametrized by two blocks consisting of a zero-padded convolution layer with $32$ filters of size $3 \times 3$ (or $3\times 3 \times 3$ for CBCT) followed by a PReLu non-linearity. The zero-padding ensures that the input and output sizes are the same. The weights of the convolutions were initialized using the Kaiming initialization. The parameter $\alpha$ was chosen to be $0.1$. See Figure~\ref{fig:lsirt} for a graphical overview and the algorithm description in Algorithm~\ref{alg:sirt}.
The network was trained with batch size $N_{b} = 8$. During the first $N_{s}$ iterations of the lSIRT algorithm, no updates of the weights $\theta$ are performed as the first iterations are dominated by the final term in \eqref{eq.learnedsirt}. We achieved good results with $N_{s} = 50$, and kept this value in all our experiments. The total number of iterations of lSIRT was set to $N_{\text{tot}} = 100$. To keep the gradients across iterations small, and to limit the computational cost, we randomly replace only single elements per batch. We do this in such a way that the average number of lSIRT iterations that is applied to an image in the batch is approximately $N_{\text{tot}}$ and each image in the batch is used approximately $N_{\text{tot}} - N_{s}$ times to compute gradient updates. 

The mean square error was used as a loss function, where we applied a different weight factor $\omega$ to the different output channels $\gamma_i$ of the neural network. Subsequently a logarithm was applied to limit the loss values in early iterations. In particular, the loss function $\mathcal L$ used is:
\begin{equation}
    \label{eq:loss-function}
    \mathcal{L}(\gamma, x, t) = \mathbf{E} \log( \|\gamma_0 - t\|_{L_2}^2 + \omega \|\gamma_1 - (t - x)\|_{L^2}^2),
\end{equation}
After some initial experimentation, we settled with $\omega = 0.04$.

We trained the network $g_\theta$ using the Adam optimizer \cite{Kingma15} with parameters $\beta_0 = 0.9$ and $\beta_1 = 0.99$.
The network was trained for a total of $N_{\text{iter}} = 80 000$ iterations with a learning rate which was set at $2 \cdot 10^{-4}$ for the first $40000$ iterations, decreasing to $5 \cdot 10^{-5}$ for the next $20000$ iterations and finally decreasing $0$ in the last $20000$ iterations, except for the $256^3$ voxel 3D models, which were trained for $50000$ iterations, with a learning rate decreasing from $10^{-4}$ to $0$.
Additionally, to reduce memory usage, we trained the $256^3$ voxel 3D models in a patch-based manner, by training on patches of $128^3$. As mentioned before, the ability to do this is an advantage above methods which also learn in the projection domain. 

We trained in total eight algorithms, four for 2D and 3D each. 
In 2D we trained models for the triangle and lung data, both for the low noise and high noise regime. 
3D models were trained on the ellipse data for the low, medium and high noise levels, and one model was trained for the lung data in the low noise regime. 
The networks were trained on a single Nvidia RTX2080Ti GPU. 
We implemented our algorithms using the PyTorch library (version 1.0.1). The projection operators $A$ and $A^T$ were computed using the Astra toolbox (version 1.8) \cite{VanAarle2015}.
\begin{algorithm}\label{alg:sirt}
	\caption{The training of lSIRT}
\begin{algorithmic}	
	\Function{createbatch}{$N$}
		\For{$i \in \{1,\dots,N\}$}
			\State Select an image $\chi$ from dataset.
			\State $x_i, y_i, t_i, h_i = 0, A(\chi) + \eta, \chi, 0$
		\For{$N_s \text{ iterations}$}
		\State $p_i = C  A^T R (A x_i - y_i)$
		\State $\gamma = g_{\theta}(x_i,h_i,p_i)_0$
		\State $h_i = x_i$ \Comment{Save $x_i$ for next iteration}
		\State $x_i = (1-\alpha x_i) + \alpha \gamma + p $
		\EndFor
		\EndFor
		\State \textbf{return} $(x_i)_i, (y_i)_i, (t_i)_i$
	\EndFunction \\
	
	\State $x,y,t =$ \Call{createbatch}{$N_b$} \Comment{Initialize first batch}
	\For{$N_{it} \text{ iterations}$}
	\If{$\operatorname{uniform}(0,1) < N_{b}/(N_{\text{tot}} - N_s)$} 
	\State $\text{Select random integer } i \in [1, N_b]$.
		\State $x_i, y_i, t_i = \text{\Call{createbatch}{$1$}}$
		\State $h_i = 0$
	\EndIf
	\State $L = 0$  \Comment(Initialize loss)
	\For{$i \in \{1,\dots,N_{b}\}$}
	\State $p_i = C  A^T R (A x_i - y_i)$
		\State $\gamma =  g_{\theta}(x_i,h_i,p_i)$
		\State $h_i = x_i$
		\State $x_i = (1 - \alpha) x_i + \alpha \gamma_0 + p_i $
		\State $L = L + \mathcal{L}(\gamma, x_i, t_i)$
		\State Backpropagate $L$ w.r.t. $\theta$.
	\EndFor
	\State Update weights $\theta$ and learning rate.
	\EndFor
\end{algorithmic}
\end{algorithm}

\section{Results for simulated projections}\label{sec:lsirt-star}
We compare the performance of the lSIRT algorithm both with classical (i.e., non-deep learning) and deep learning based methods. In particular we compare with the classical FBP and SIRT algorithms, a learned post-processor based on a U-net and the Learned Primal Dual algorithm (LPD) \cite{Adler18}. We also study the effect of not using the ``history" \eqref{eq.genh} and refer to this algorithm as lSIRT*. The specifics of the deep learning based methods are described in this section.

\subsubsection{lSIRT*}
The lSIRT* algorithm is basically the lSIRT algorithm where $g_\theta$ only receives information from $x^{(k + 1)}$ and not from its neighbors through \eqref{eq.genh}. This requires a change in the loss function \eqref{eq:loss-function} as well, by dropping the final term (e.g. $\omega = 0$). This is also denoted in Figure~\ref{fig:lsirt} where the dashed lines are omitted in contrast to lSIRT.

\subsubsection{Learned primal dual}
The LPD algorithm \cite{Adler18} is included into the comparison in 2D. There is no methodological limitation why the method cannot be applied in 3D, but the memory requirements quickly surpass the available GPU memory limiting the comparison to 2D.

Next to the algorithm as described in the paper (henceforth: \emph{LPDorig}), we also included another version where we share the weights between subsequent primal-dual blocks (henceforth: \emph{LPDsame}). This in effect brings the number of learnable parameters closer to those found in lSIRT. Compared to the algorithm described in \cite{Adler18}, we make some small changes. We selected a batch size of $3$ and used a learning rate schedule which linearly decreases from $2 \cdot 10^{-4}$ to $0$ in $100 000$ iterations.

\subsubsection{SIRT + U-net post-processing}
To compare with learned post-processor, we trained a U-net to remove artifacts from the SIRT reconstruction. A U-net with depth $4$ was used, where the downsampling block consists out of two $3 \times 3$ zero-padded convolutions with the same number of filters, followed by instance normalization, a ReLu activation and a max-pooling layer with stride $2$. The upsampling path had a similar structure with the max-pooling layer replaced by a bilinear upsampling layer. The skip connections concatenated the output of the downsampling path to the corresponding upsampling path. Similarly to the other methods, learning was completed after $100 000$ iterations, using the Adam optimizer and a learning rate that linearly decreases from $10^{-3}$ to $0$.

In Table~\ref{tab:training}, we provide more information on the number of parameters of the neural networks used in the learned models, and an estimate of the GPU memory usage.
\begin{table}
\caption{Training requirements}
\centering
\begin{tabular}{l | c c }
	Model & Number of & {GPU}\\
     & parameters & memory\\
\hline
LPDsame & 26k & 2.8GB \\
LPDorig  & 258k & 2.8GB \\
U-net & 13M  & 2.2GB \\
lSIRT* & 9.9k & 1.1GB\\
lSIRT & 11k & 1.1GB \\
lSIRT3D@$128^3$ & 32k & 3.1GB \\
lSIRT3D@$256^3$ & 32k & 7.6GB \\
\hline
\end{tabular}\\
Number of parameters and estimated amount of GPU memory used for a batch size of $1$.
\label{tab:training}
\end{table}

\subsection{Trained models}

\subsubsection{2D models}
\begin{figure}
	\centering
	\subfigure[truth]                         {\includegraphics[width = 1.1in]{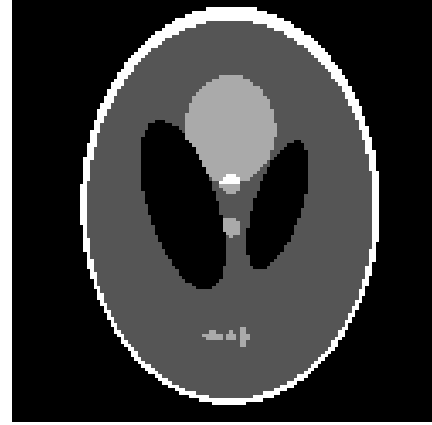}}
	\subfigure[FBP]                           {\includegraphics[width = 1.1in]{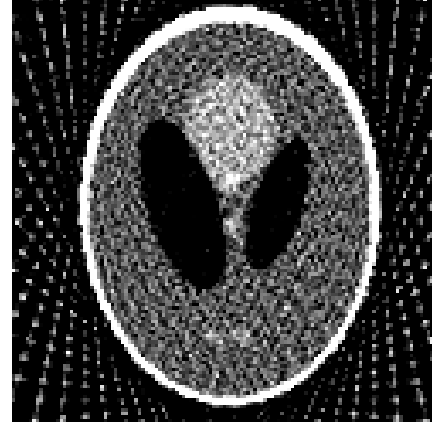}}
	\subfigure[SIRT]                          {\includegraphics[width = 1.1in]{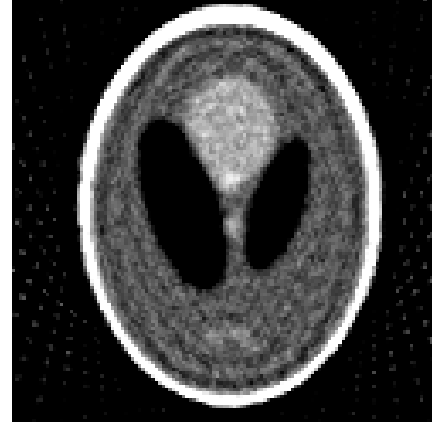}}\\
	\subfigure[U-net]                         {\includegraphics[width = 1.1in]{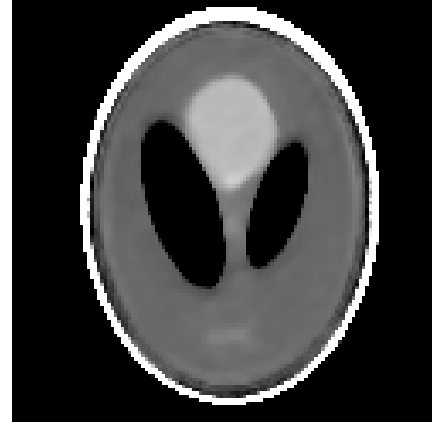}}
	\subfigure[Primal dual same ]             {\includegraphics[width = 1.1in]{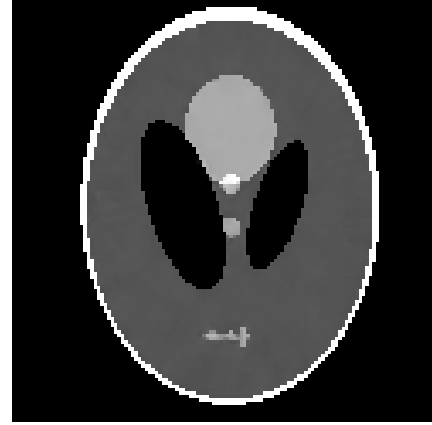}}
	\subfigure[learned SIRT]                  {\includegraphics[width = 1.1in]{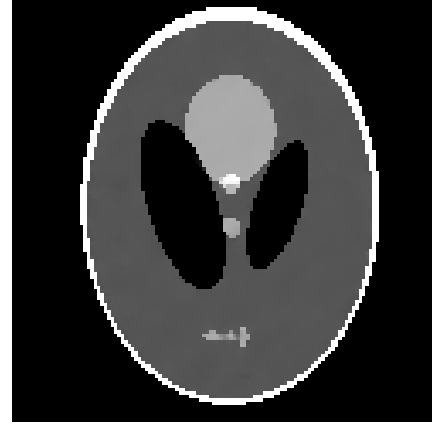}}\\

	\caption{Output of the models that are trained on low-noise triangles, evaluated on a low-noise Shepp Logan phantom with window $[100, 400]$HU.}
	\label{fig:sl}
\end{figure}

In Table~\ref{tab:triangles} the performance of the 2D model trained on triangles, using the same projection geometry as during training, for multiple noise levels is given. 
Figure~\ref{fig:sl} provides examples of the reconstruction of the Shepp Logan phantom where the learned algorithms are trained on the triangle data.
The LPDsame, LPDorig and lSIRT all achieve comparable results and outperform the U-net post-processing.

\begin{figure*}
	\centering
	\subfigure{\includegraphics[width = 0.3\linewidth]{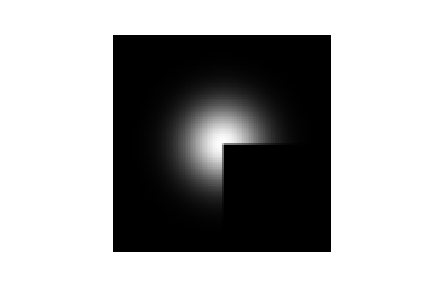}}
	\subfigure{\includegraphics[width = 0.3\linewidth]{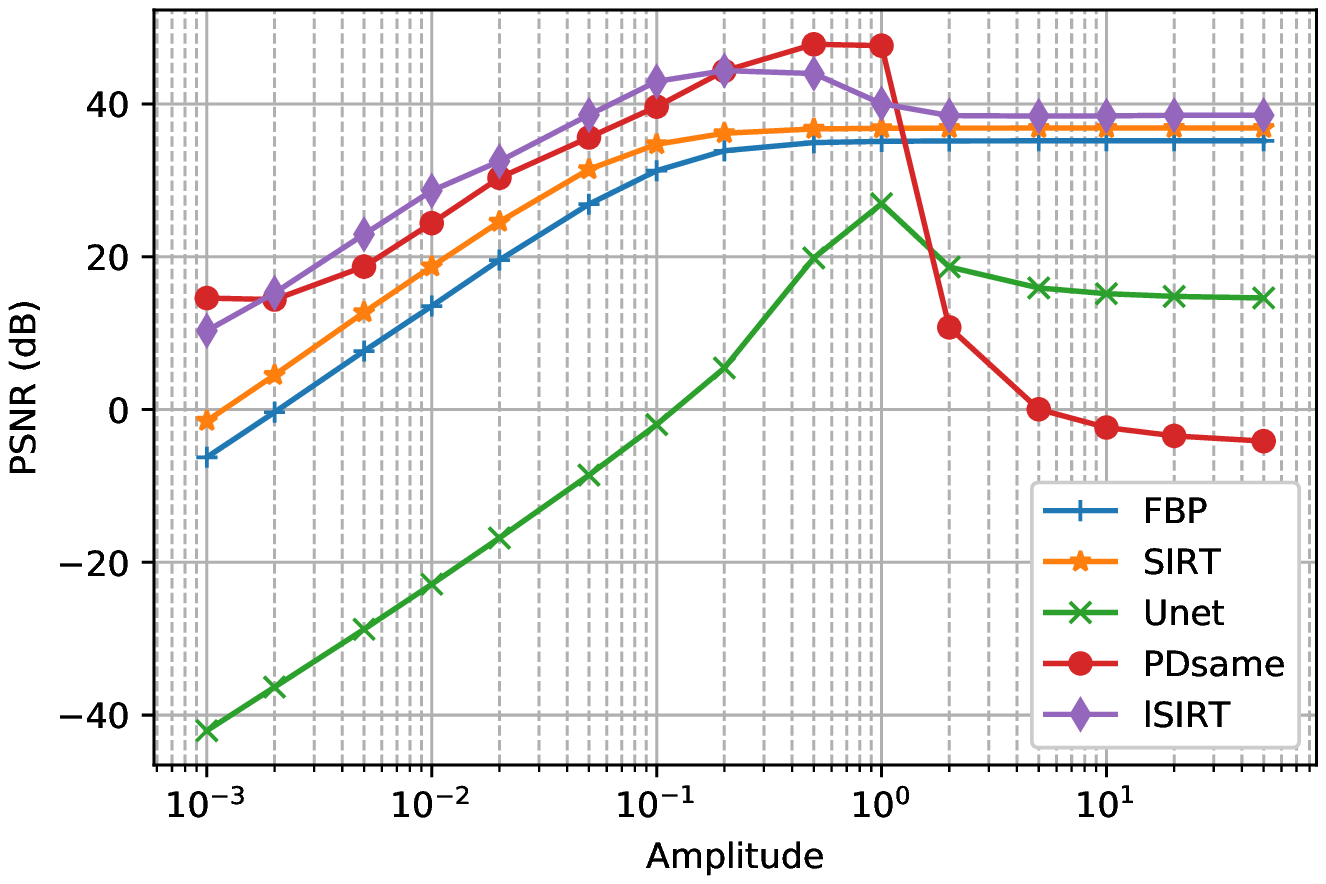}}
	\subfigure{\includegraphics[width = 0.3 \linewidth]{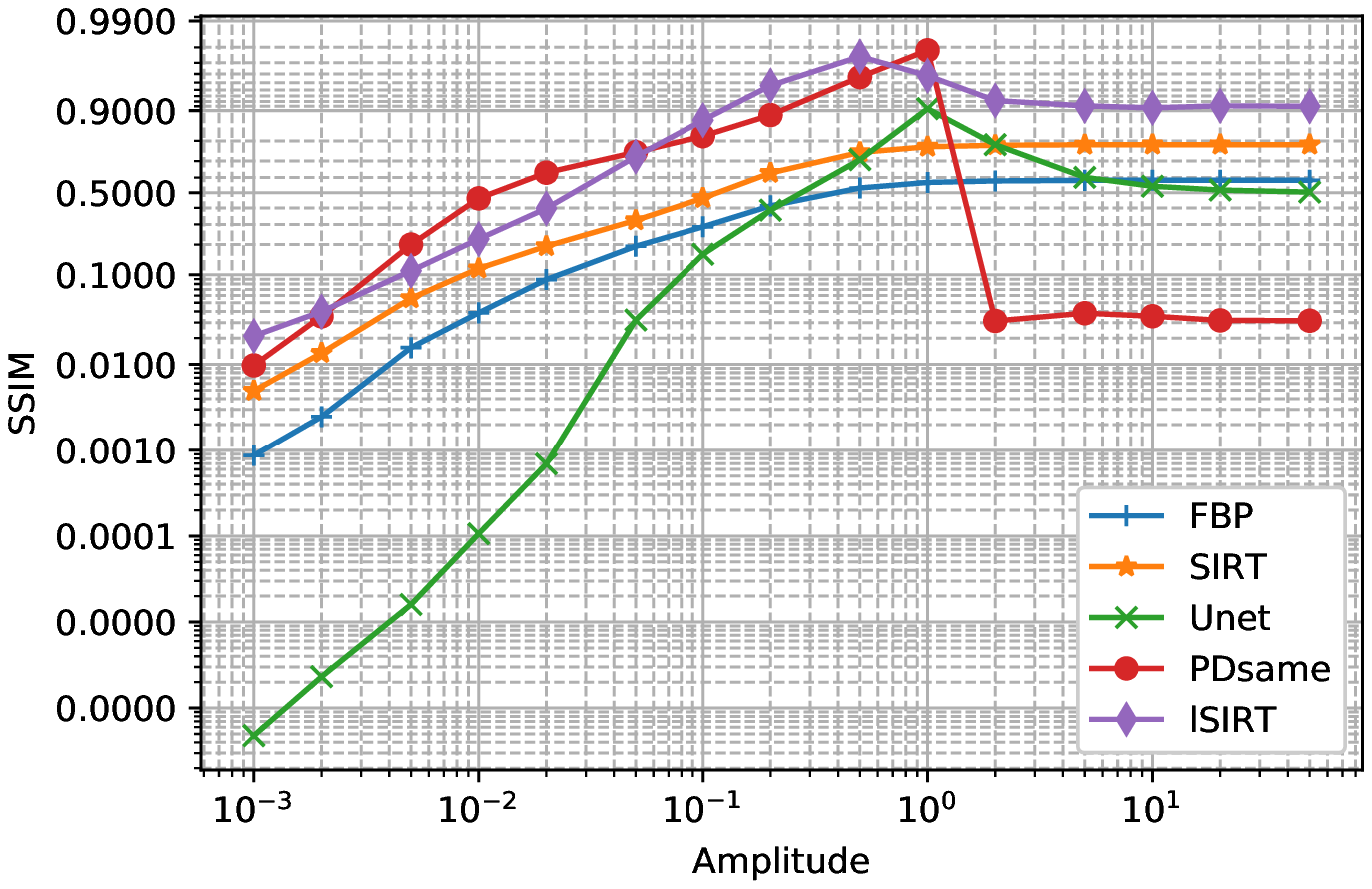}}
	\caption{(a) Gaussian phantom with a square left out. (b,c) The PSNR and SSIM for different amplitudes of the Gaussian phantom.}
	\label{fig:strange}
\end{figure*}

To evaluate the stability we perform two additional tests: (i) we use lSIRT* (lSIRT with the dotted arrows removed in Figure~\ref{fig:lsirt}).  In all experiments we show superiority of the lSIRT algorithm, at minimal extra computational cost; (ii) to investigate the generalizability of the model, we test the model on out-of-distribution data. For this we create a synthetic phantom consisting of a Gaussian $A \cdot \exp(-0.002(x^2+y^2))$ with a square area zeroed out, with $x$ and $y$ in millimeters. This image contains smooth gradients in contrast to the triangle set on which the model was trained. The value $A$ allows to study the stability of the algorithm for different amplitudes. An example for $A = 1$ is given in Figure~\ref{fig:strange}(a). In Figure~\ref{fig:strange}(b, c) the effect of the parameter $A$ on the PSNR and SSIM is plotted. These graphs suggest that learned methods such as the LPD and the U-net post processor are much more sensitive to out-of-distribution samples. This is not unexpected as the complexity of their networks allows to learn more structure in the dataset as well.

\begin{figure}
	\centering
	\subfigure[truth]                         {\includegraphics[width = 1.1in]{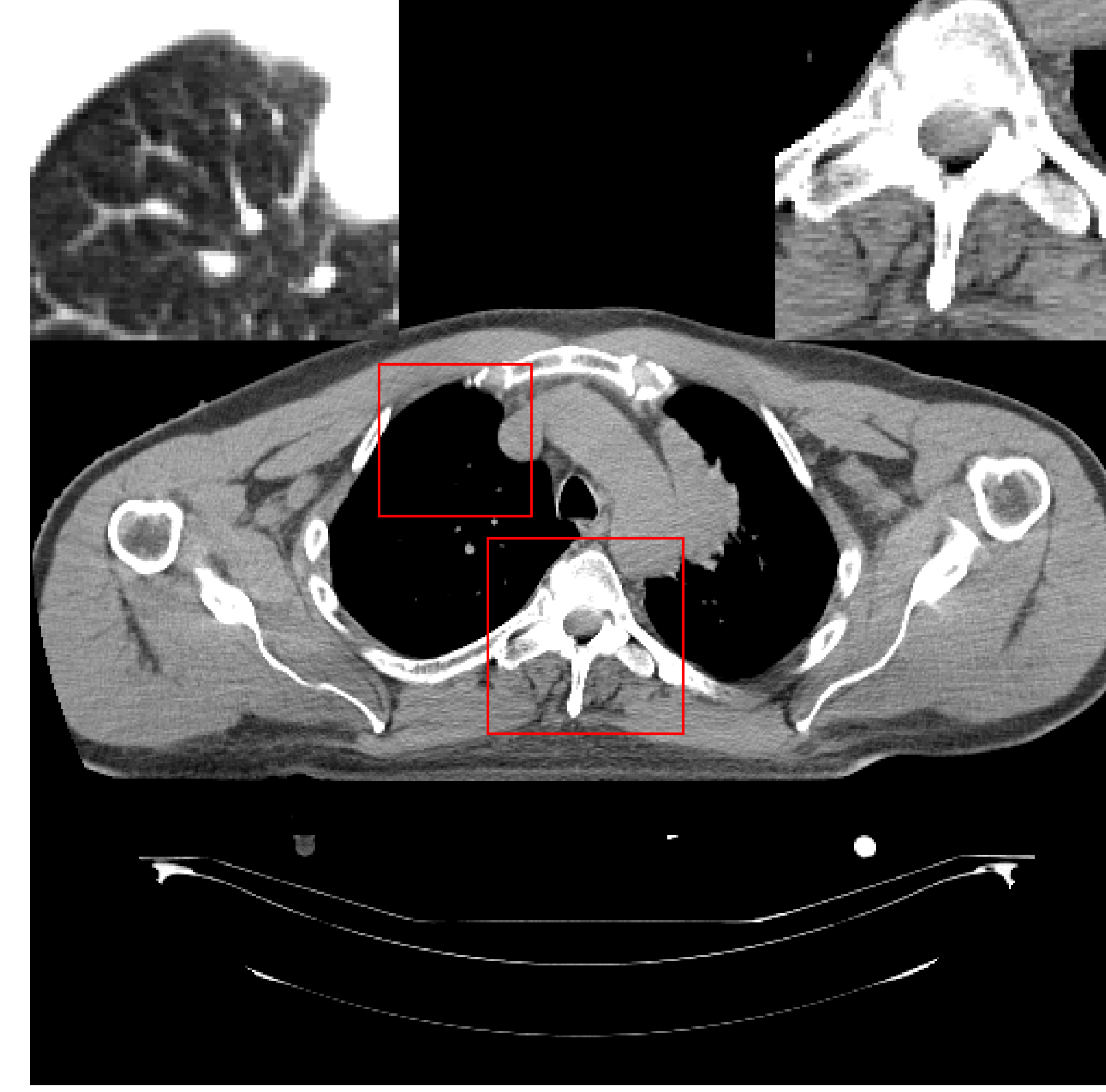}}
	\subfigure[FBP]                           {\includegraphics[width = 1.1in]{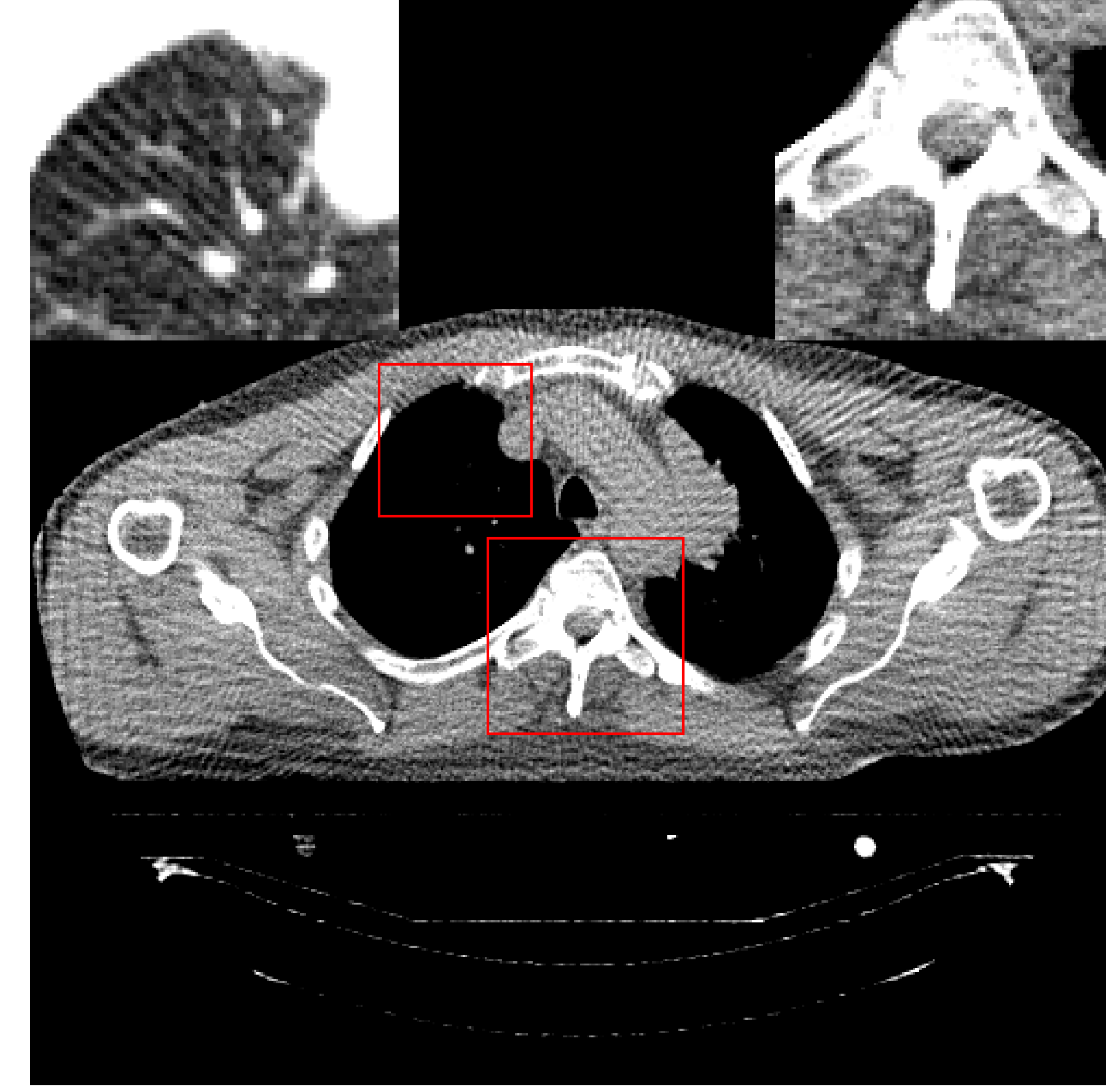}}
	\subfigure[SIRT]                          {\includegraphics[width = 1.1in]{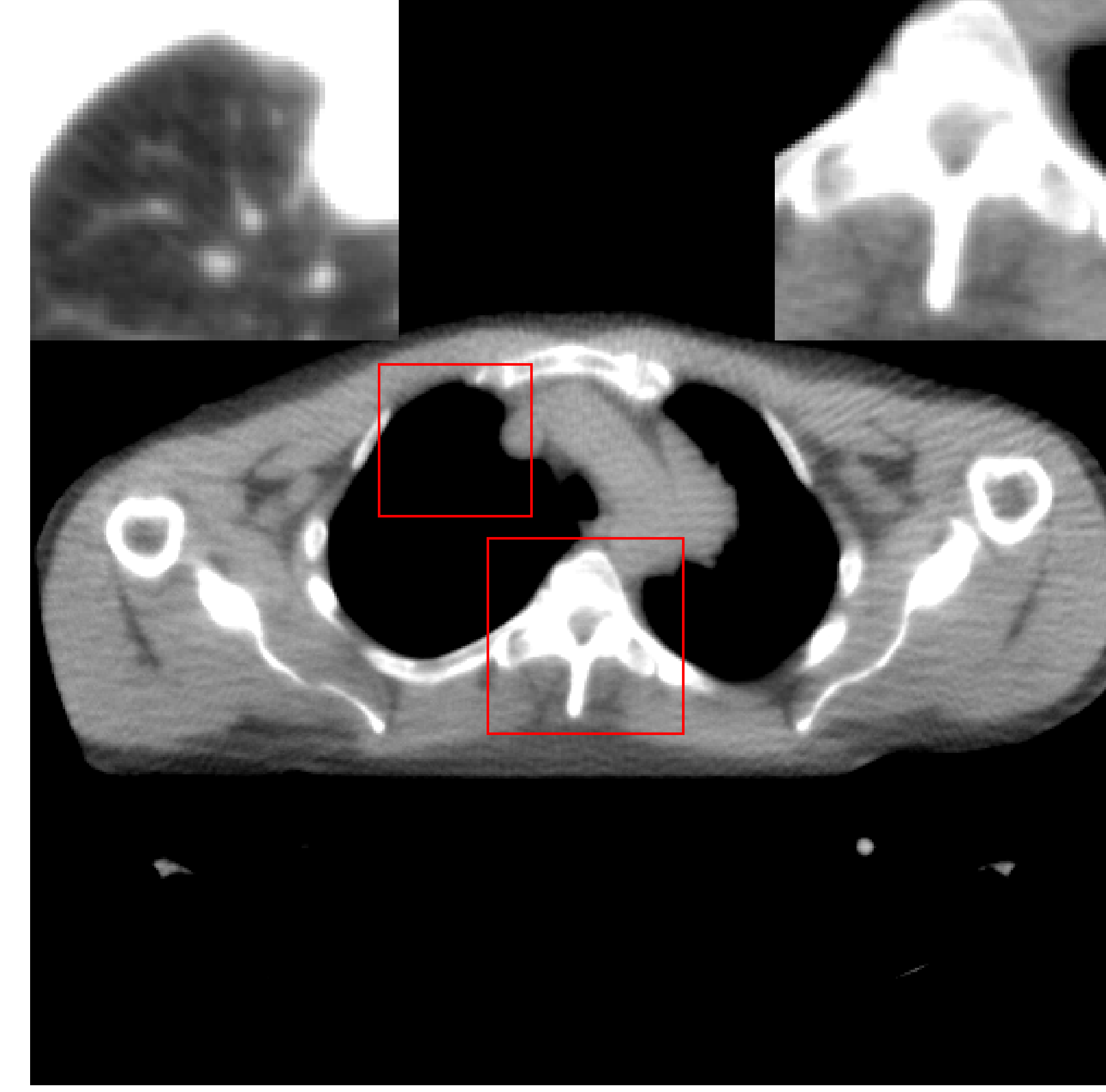}}\\
	\subfigure[U-net]                         {\includegraphics[width = 1.1in]{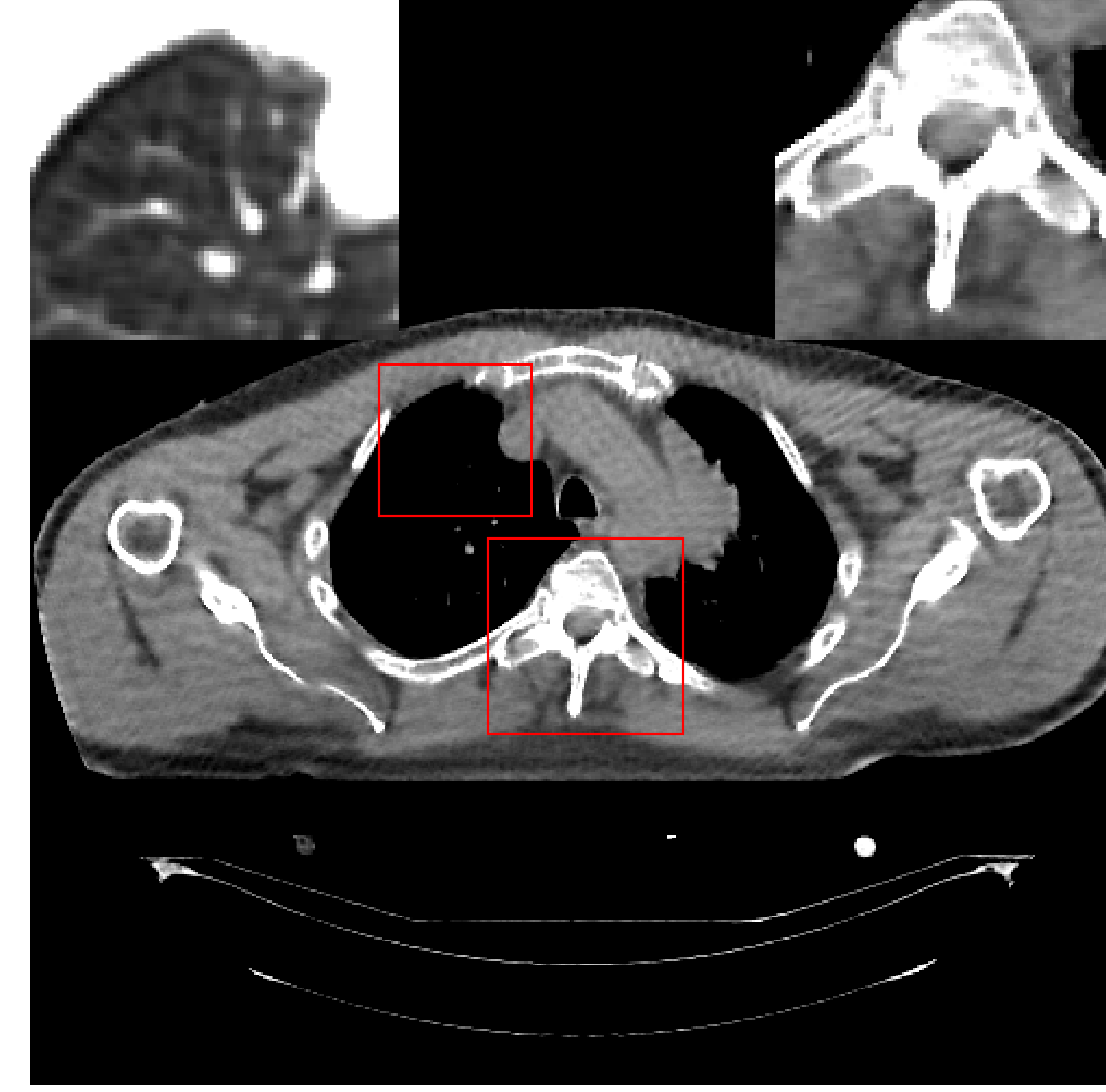}}
	\subfigure[Primal dual original ]        {\includegraphics[width = 1.1in]{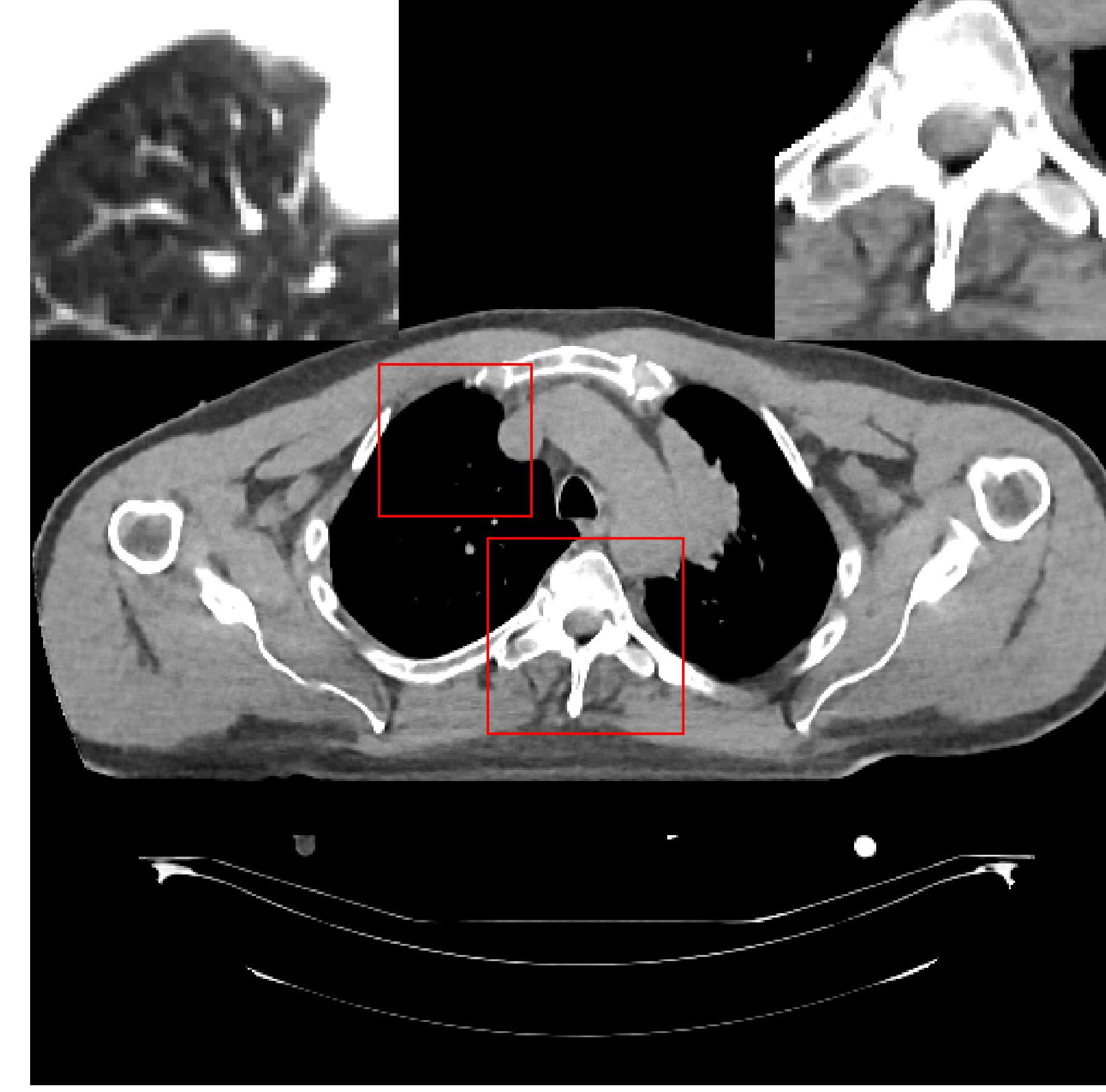}}
	\subfigure[learned SIRT]                  {\includegraphics[width = 1.1in]{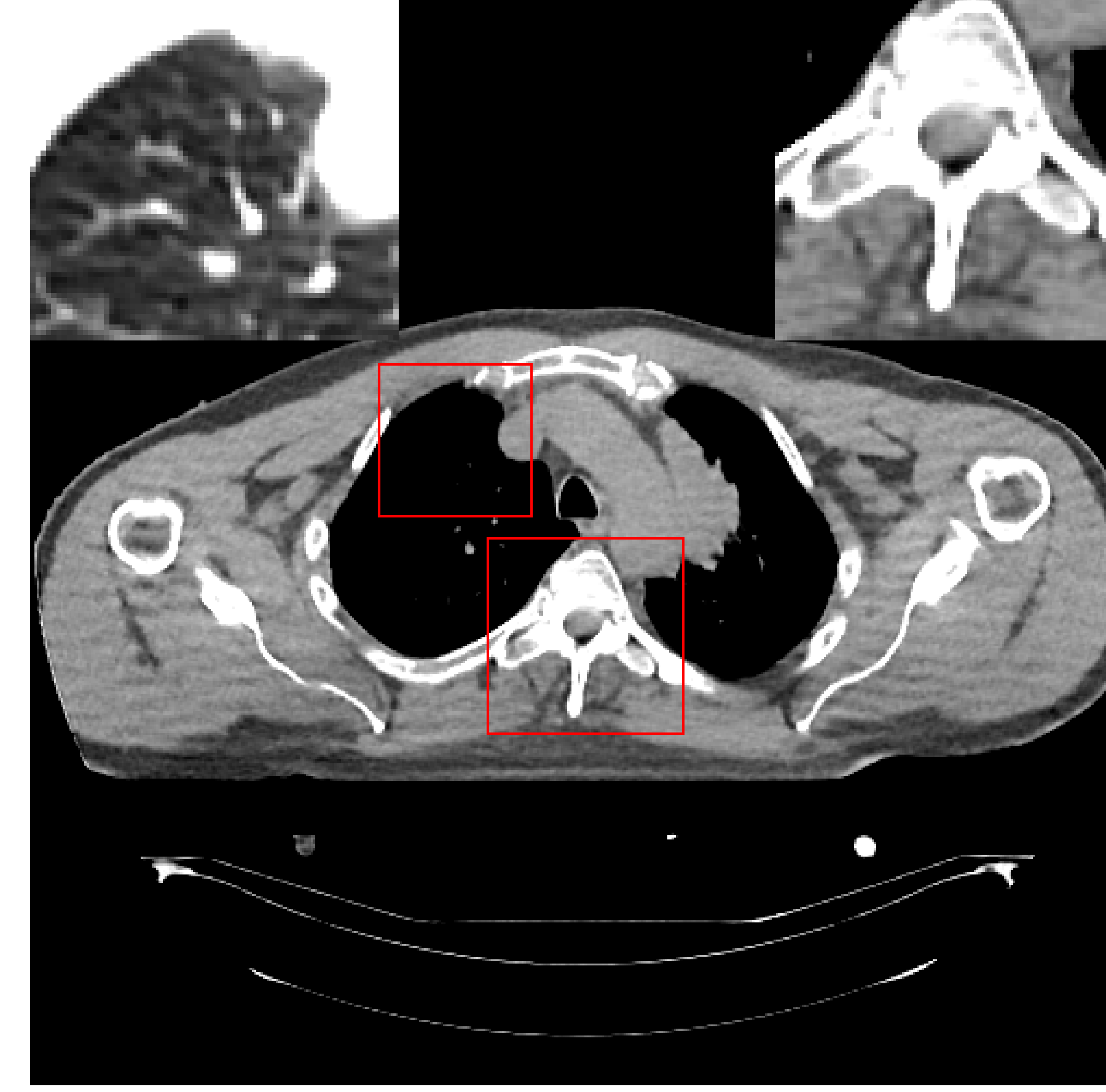}}\\

	\caption{Output of the models on a slice of the validation set with low-noise conditions, with window levels of $[-200,200] \text{HU}$, $[-900,-400] \text{HU}$, and $[-200,200] \text{HU}$ for the main image, left insert and right insert, respectively.}
	\label{fig:patient}
\end{figure}

\begin{figure}
	\centering
	\subfigure[truth]                         {\includegraphics[width = 1.1in]{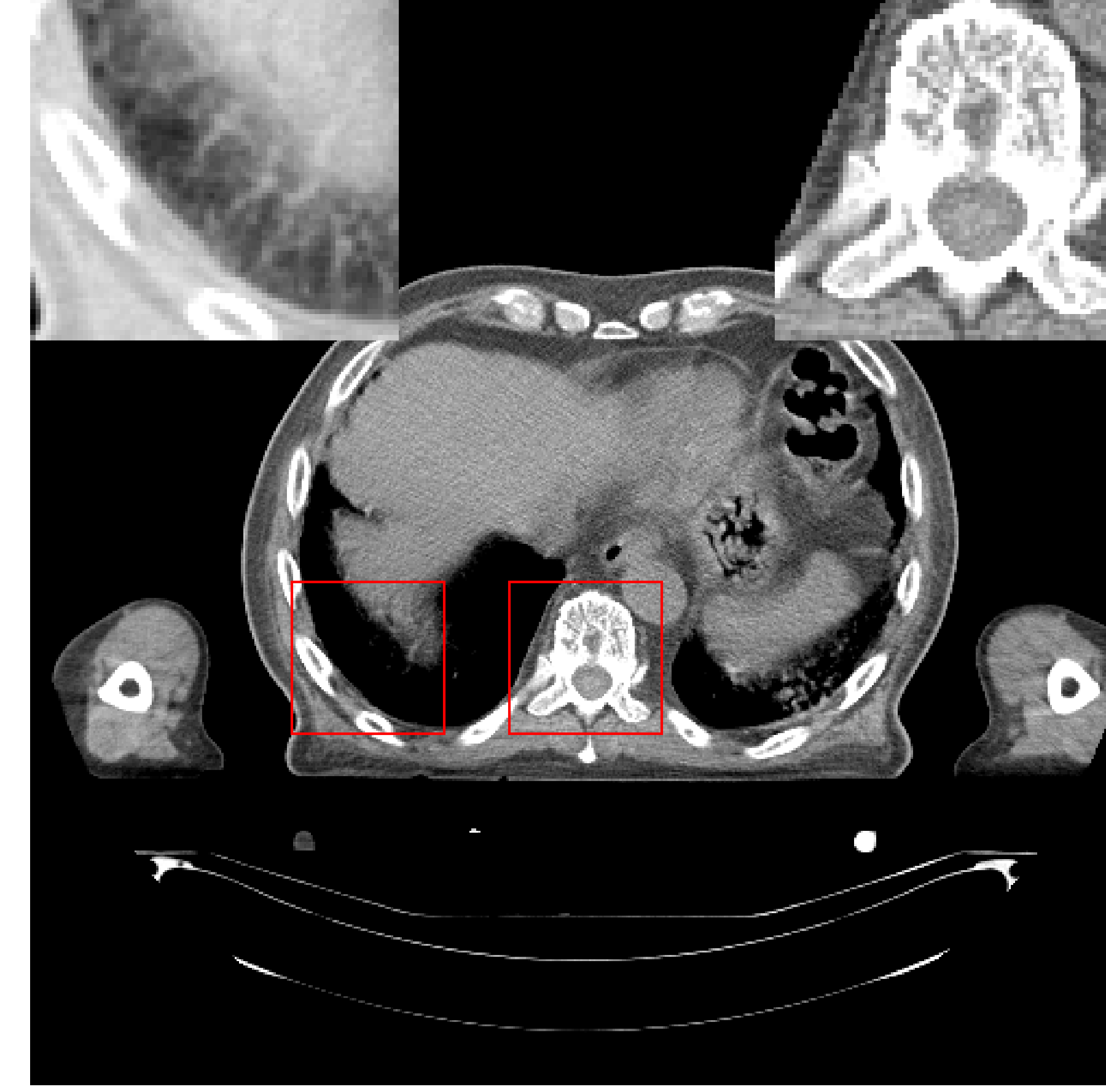}}
	\subfigure[FBP]                           {\includegraphics[width = 1.1in]{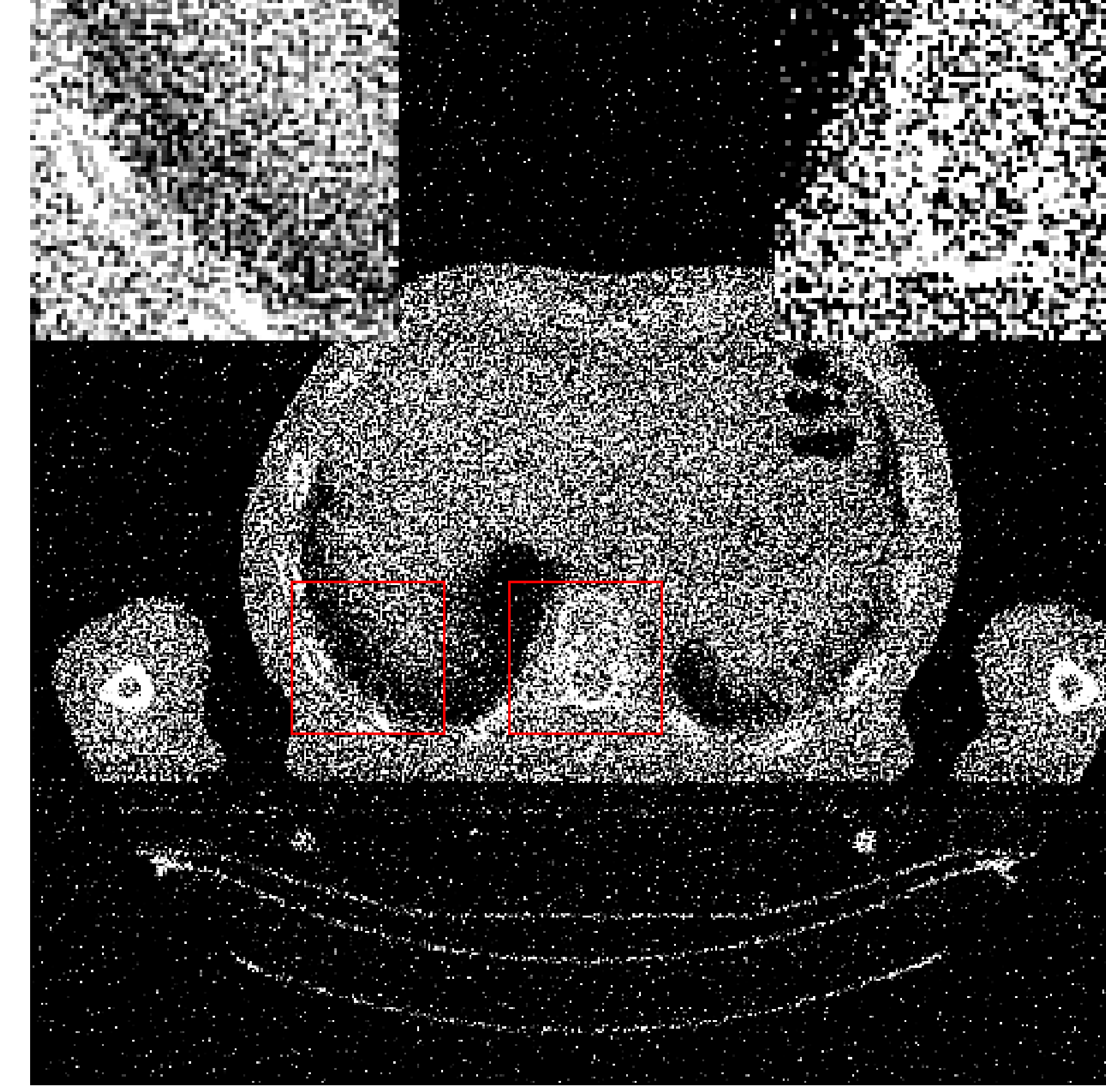}}
	\subfigure[SIRT]                          {\includegraphics[width = 1.1in]{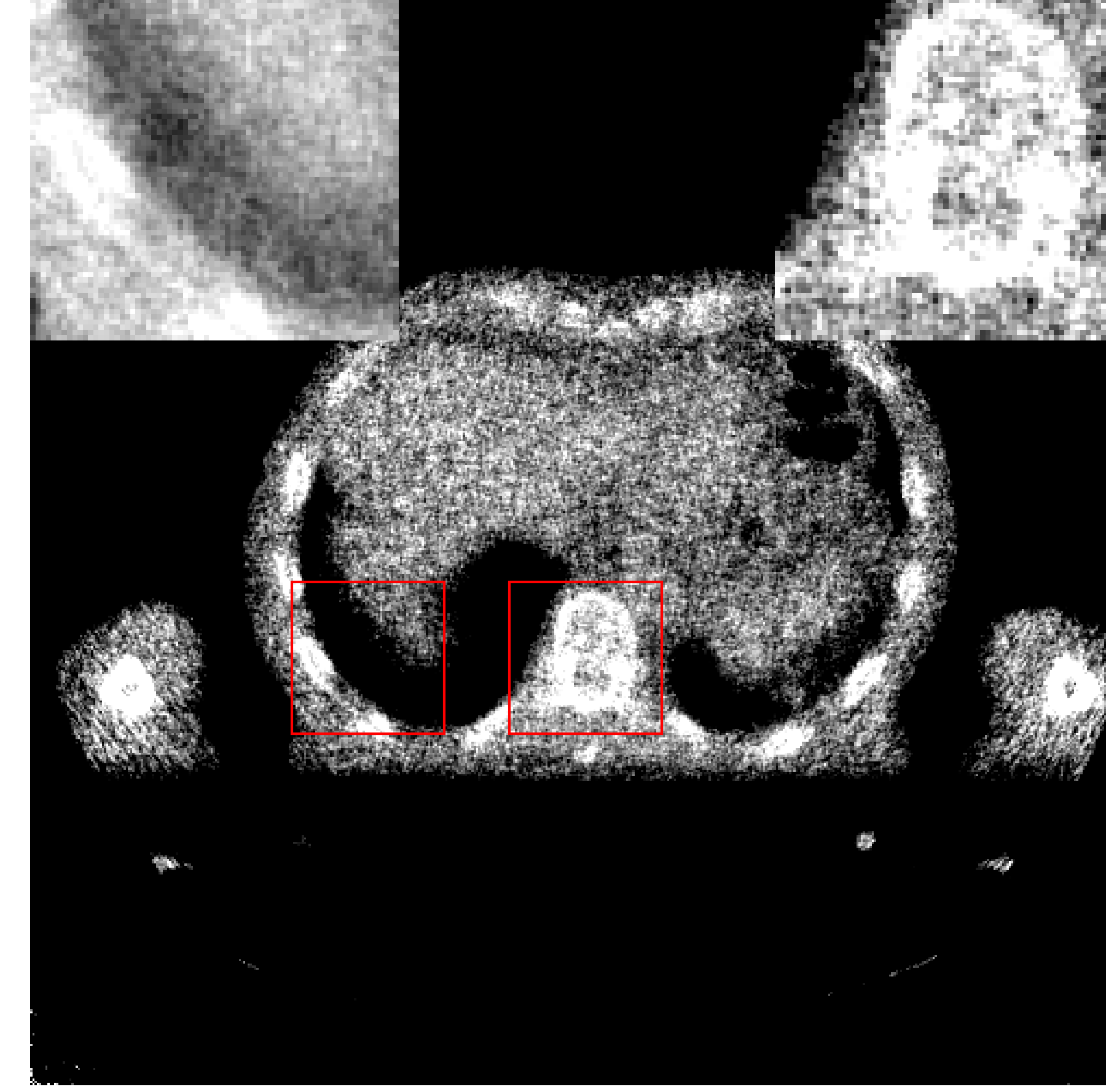}}\\
	\subfigure[U-net]                         {\includegraphics[width = 1.1in]{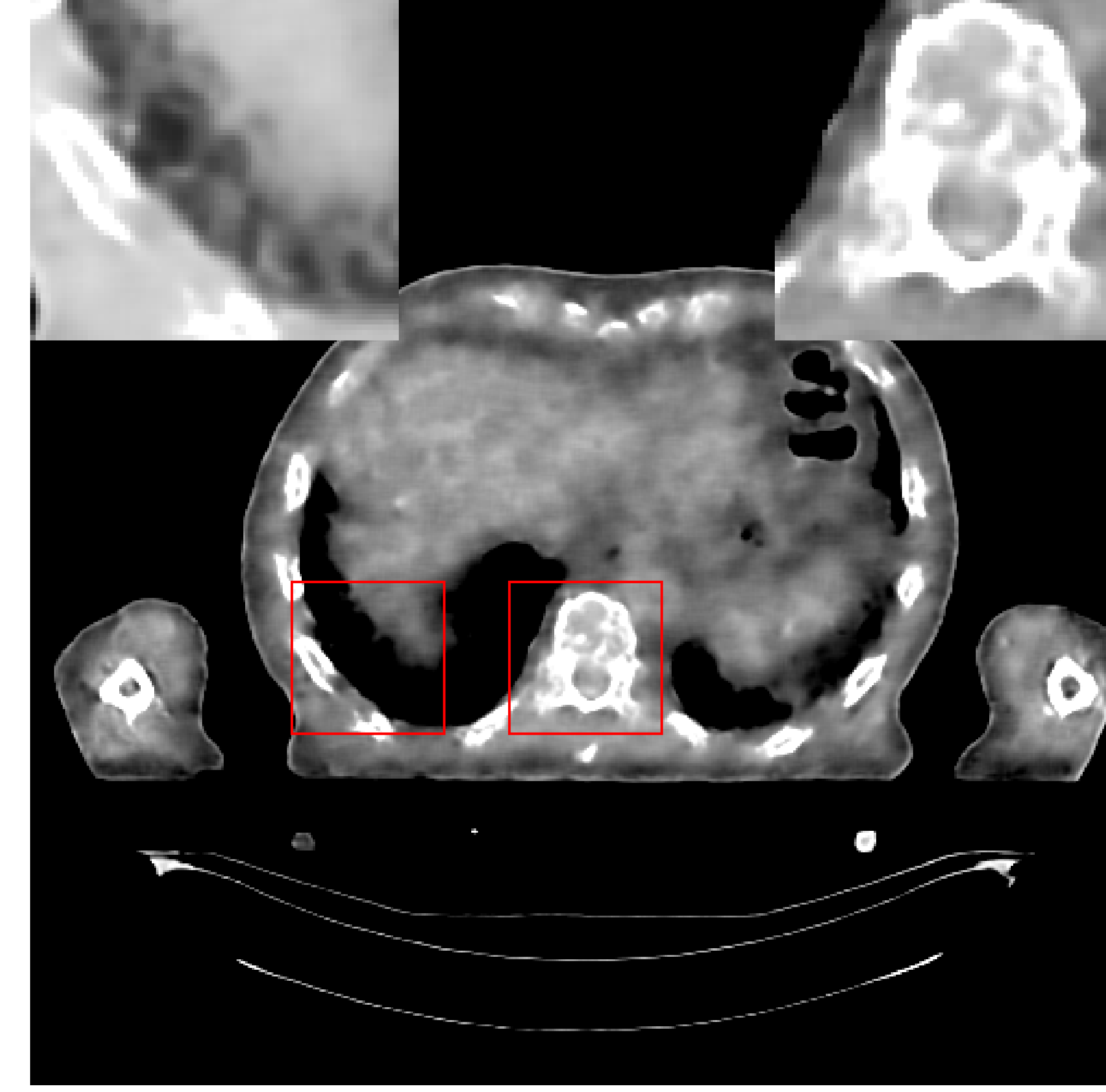}}
	\subfigure[Primal dual originial]        {\includegraphics[width = 1.1in]{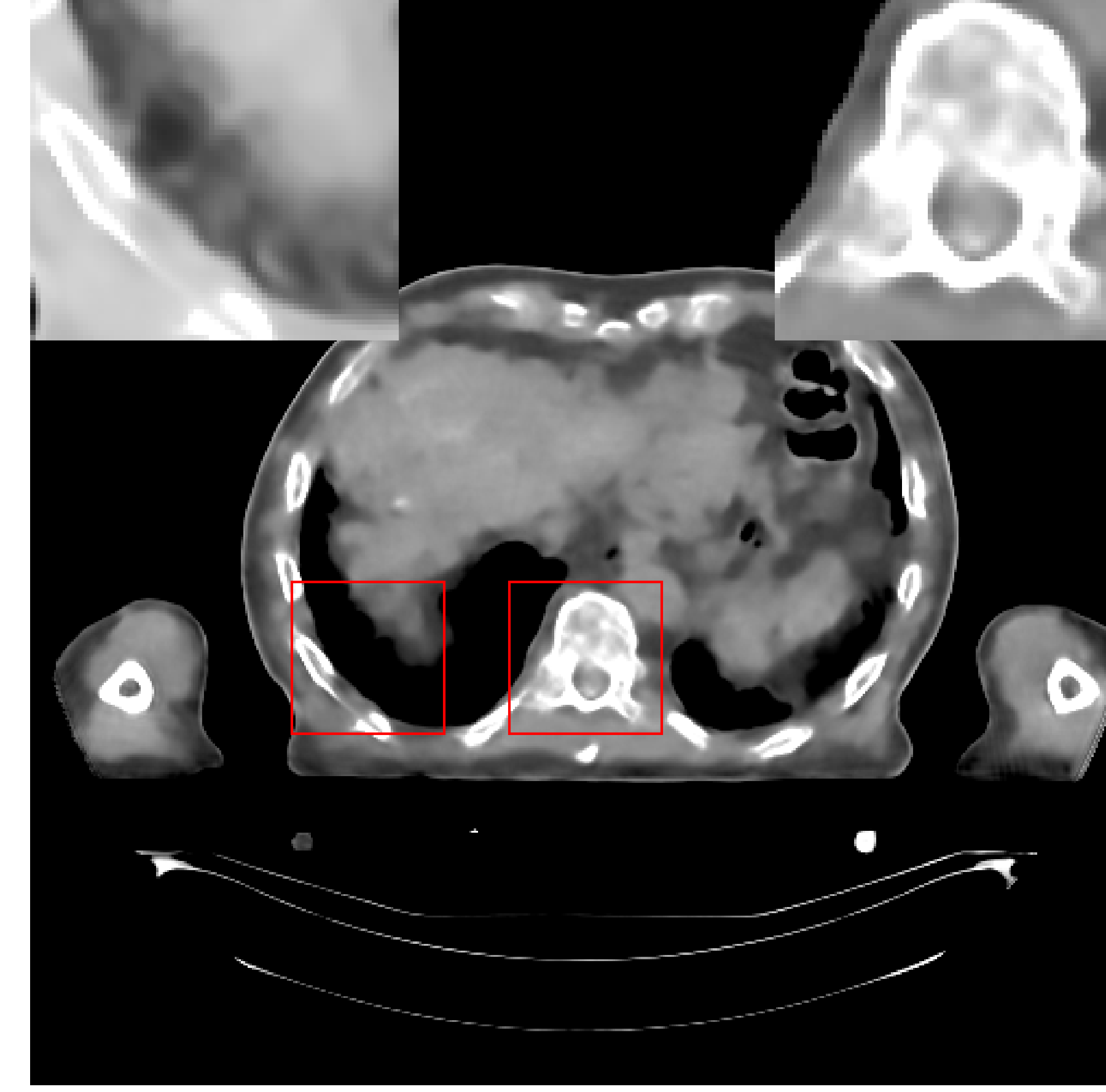}}
 	\subfigure[learned SIRT]                  {\includegraphics[width = 1.1in]{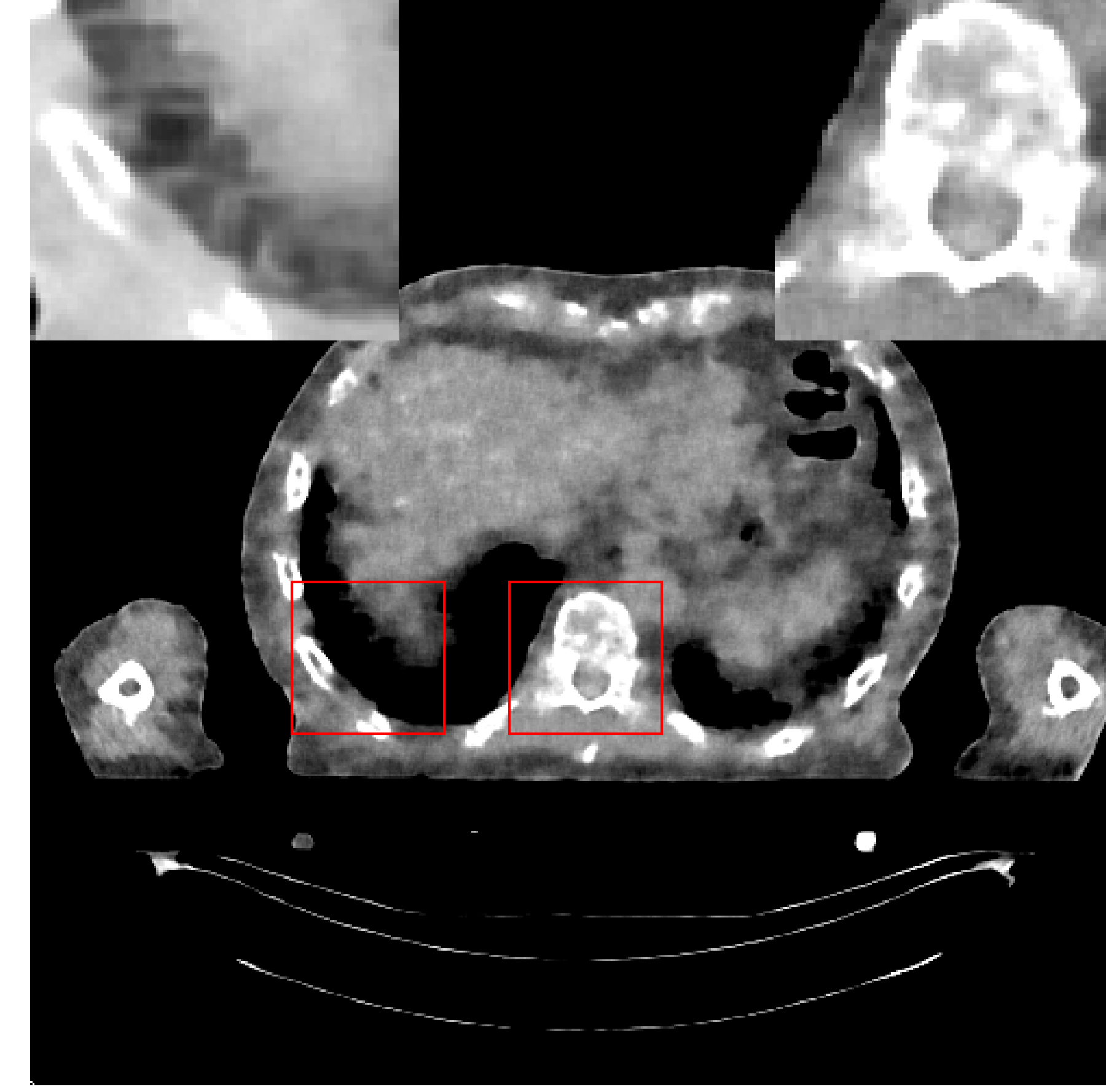}}\\

	\caption{Output of the models on a slice of the validation set with high-noise conditions, with window levels of $[-200,200] \text{HU}$, $[-900,200] \text{HU}$, and $[-200,200] \text{HU}$ for the main image, left insert and right insert, respectively.}
	\label{fig:patienthn}
\end{figure}

\begin{table}
	\centering
	\caption{Reconstruction quality for models trained on 2D triangles}
\begin{tabular}{l| l| l | c c c}
&&Experiment & PSNR & SSIM  & Runtime\\
&& & (dB) &  & (ms)\\
	\hline
\parbox[t]{2mm}{\multirow{14}{*}{\rotatebox[origin=c]{90}{Triangle phantoms}}}&
	\parbox[t]{2mm}{\multirow{7}{*}{\rotatebox[origin=c]{90}{Low Noise}}} 
	 &	FBP & $24.1$ & $0.420$ & $8.9$\\
	&&	SIRT & $26.2$ & $0.602$ & $332$\\
	&&	U-net & $34.2$ & $0.905$ & $502$\\
	&&	LPDorig & $55.2$ & $0.9987$ & $49$\\
	&&	LPDsame & $\mathbf{59.5}$ & $\mathbf{0.99937}$ & $48$\\
	&&	lSIRT* & $ 30.8$& $0.933$ & $361$\\
	&&	lSIRT & $52.2$ & $0.9948$ & $335$\\
	\cline{2-6}
	&
	\parbox[t]{2mm}{\multirow{7}{*}{\rotatebox[origin=c]{90}{High Noise}}} 
	 &	FBP & $20.3$ & $0.225$ & $7.5$\\
	&&	SIRT & $24.0$ & $0.356$ & $335$\\
	&&	U-net & $31.3$ & $0.899$ & $504$\\
	&&	LPDorig & $\mathbf{32.2}$& $0.932$ & $49$\\
	&&	LPDsame & $31.4$ & $0.913$ & $49$\\
	&&	lSIRT* & $29.4$ & $0.809$ & $373$\\ 
	&&	lSIRT & $\mathbf{32.2}$ & $\mathbf{0.973}$ & $368$\\
\hline
\parbox[t]{2mm}{\multirow{14}{*}{\rotatebox[origin=c]{90}{Shepp Logan phantom}}}&
	\parbox[t]{2mm}{\multirow{7}{*}{\rotatebox[origin=c]{90}{Low Noise}}} 
	&	FBP & $19.0$ & $0.500$ & $8$\\
	&&	SIRT & $20.5$ & $0.675$ & $338$\\
	&&	U-net & $28.9$ & $0.872$ & $497$ \\
	&&	LPDorig & $49.2$ & $0.9990$ & $49$\\
	&&	LPDsame & $52.2$ & $\mathbf{0.9994}$ & $47$\\
	&&	lSIRT* & $28.4$ & $0.978$ & $363$\\
	&&	lSIRT & $\mathbf{52.4}$ & $0.99935$ & $364$\\
	\cline{2-6}
	&
	\parbox[t]{2mm}{\multirow{7}{*}{\rotatebox[origin=c]{90}{High Noise}}} 
	&	FBP & $12.6$ & $0.209$ & $8$\\
	&&	SIRT & $16.5$ & $0.372$ & $335$\\
	&&	U-net & $23.5$ & $0.916$ & $502$\\
	&&	LPDorig & $24.3$ & $\mathbf{0.929}$ & $50$\\
	&&	LPDsame & $24.5$ & $0.921$ & $49$\\
	&&	lSIRT* & $23.5$ & $0.837$ & $361$\\
	&&	lSIRT & $\mathbf{25.3}$ & $0.857$ & $343$\\
\hline
\end{tabular}
    \vspace{5mm}
    \newline
	For the triangles, this is an average over 100 different triangle-images. For the Shepp Logan phantom, the image is kept the same, but an average is taken over 100 noise samples.
\label{tab:triangles}
\end{table}
Next to triangles, we also train on the lung data. Some examples are given in Figures~\ref{fig:patient} and Figure~\ref{fig:patienthn}. 
The best visual results are obtained with the lSIRT and the LPD algorithms. This is also reflected in the PSNR and SSIM, see Table~\ref{tab:patient} for details. Note that the SSIM has been computed based on a 2000HU data range, which well approximates the range in most examples.
In the high noise regime, the lSIRT algorithm performs worse on PSNR, yet has a superior SSIM compared to the other learned methods. 

\begin{figure}
	\centering
	\subfigure[truth            ]		{\includegraphics[width = 1.1in]{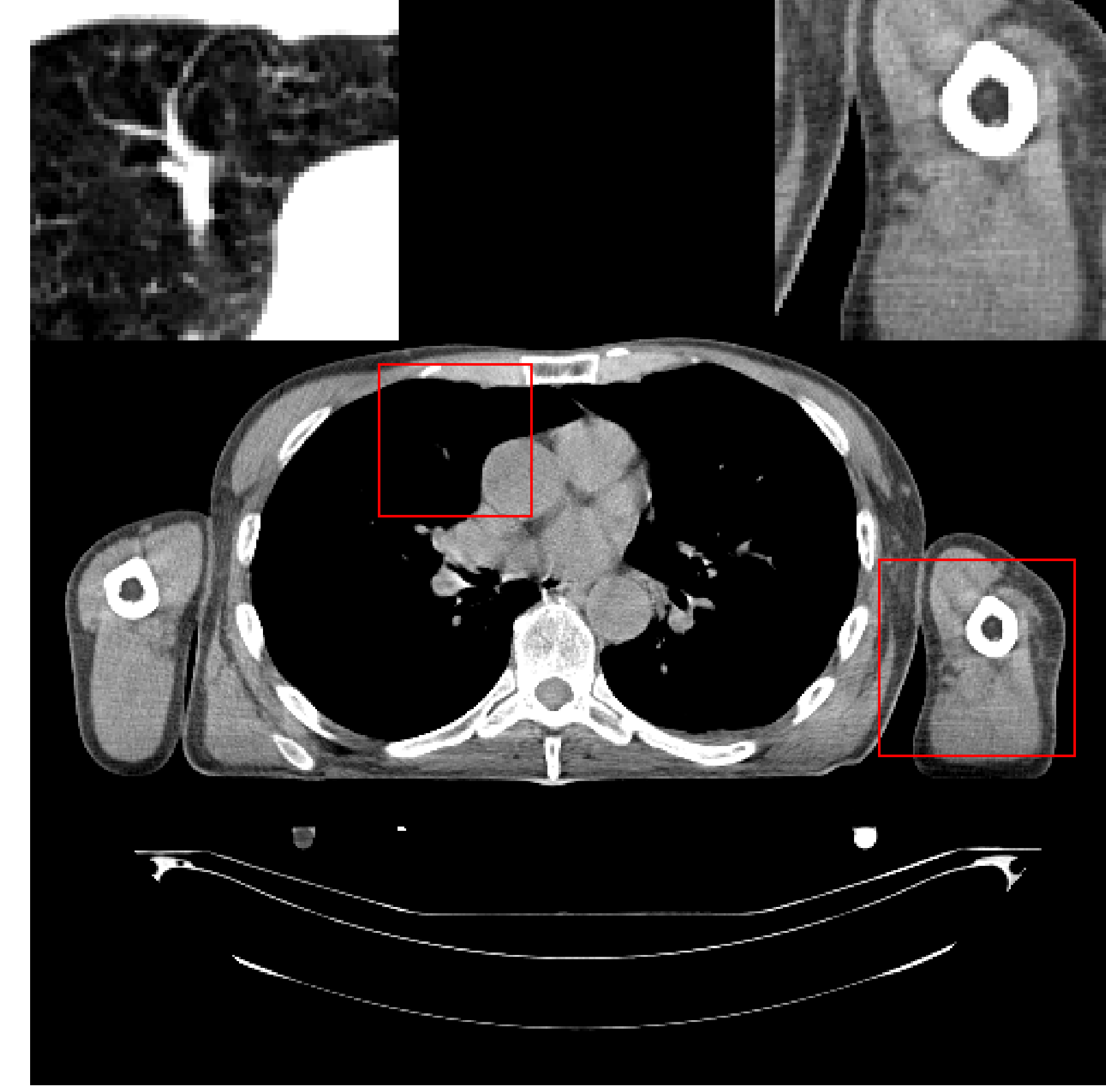}}
	\subfigure[lSIRT, 10 iterations]	{\includegraphics[width = 1.1in]{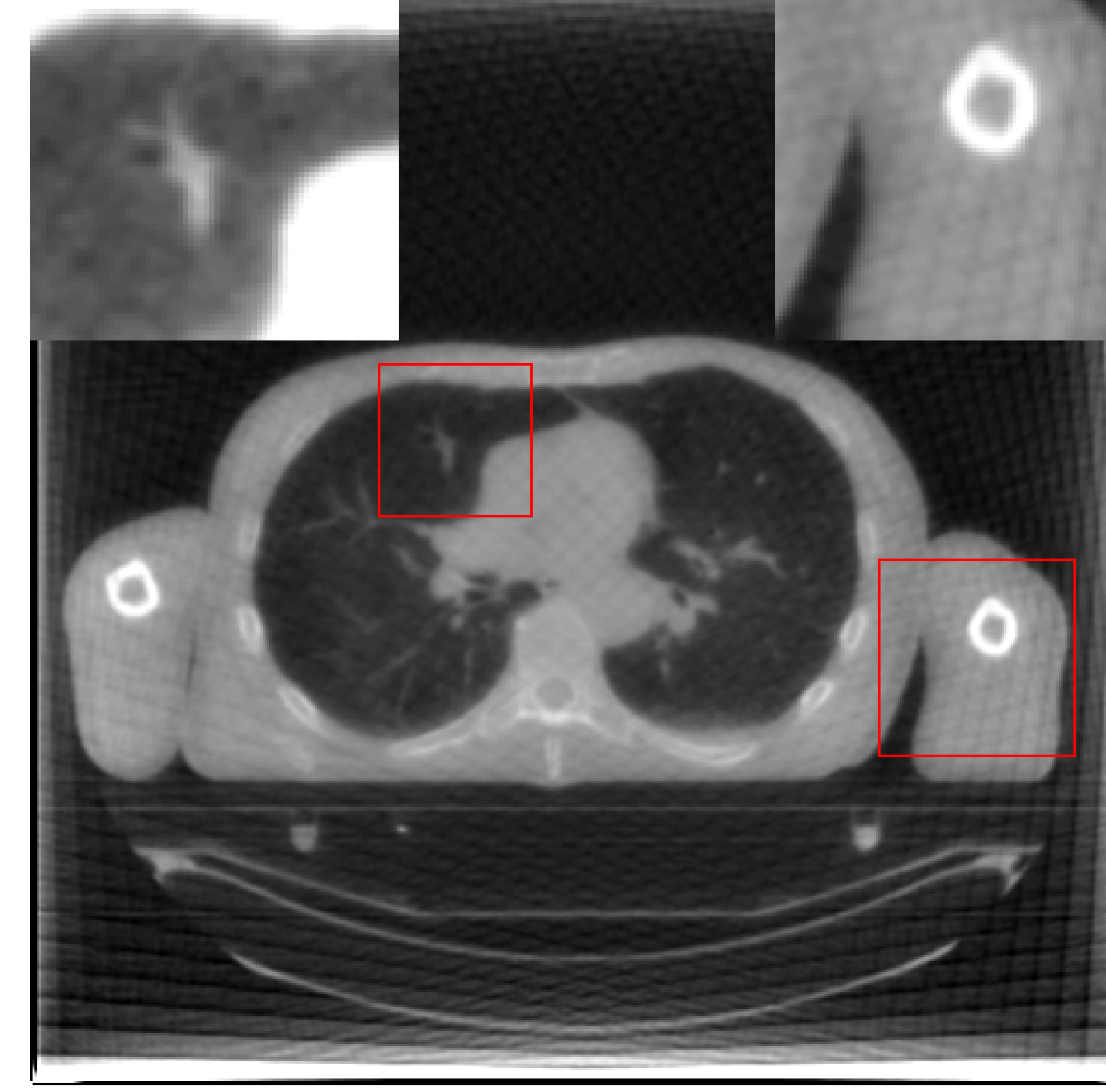}}
	\subfigure[lSIRT, 20 iterations]	{\includegraphics[width = 1.1in]{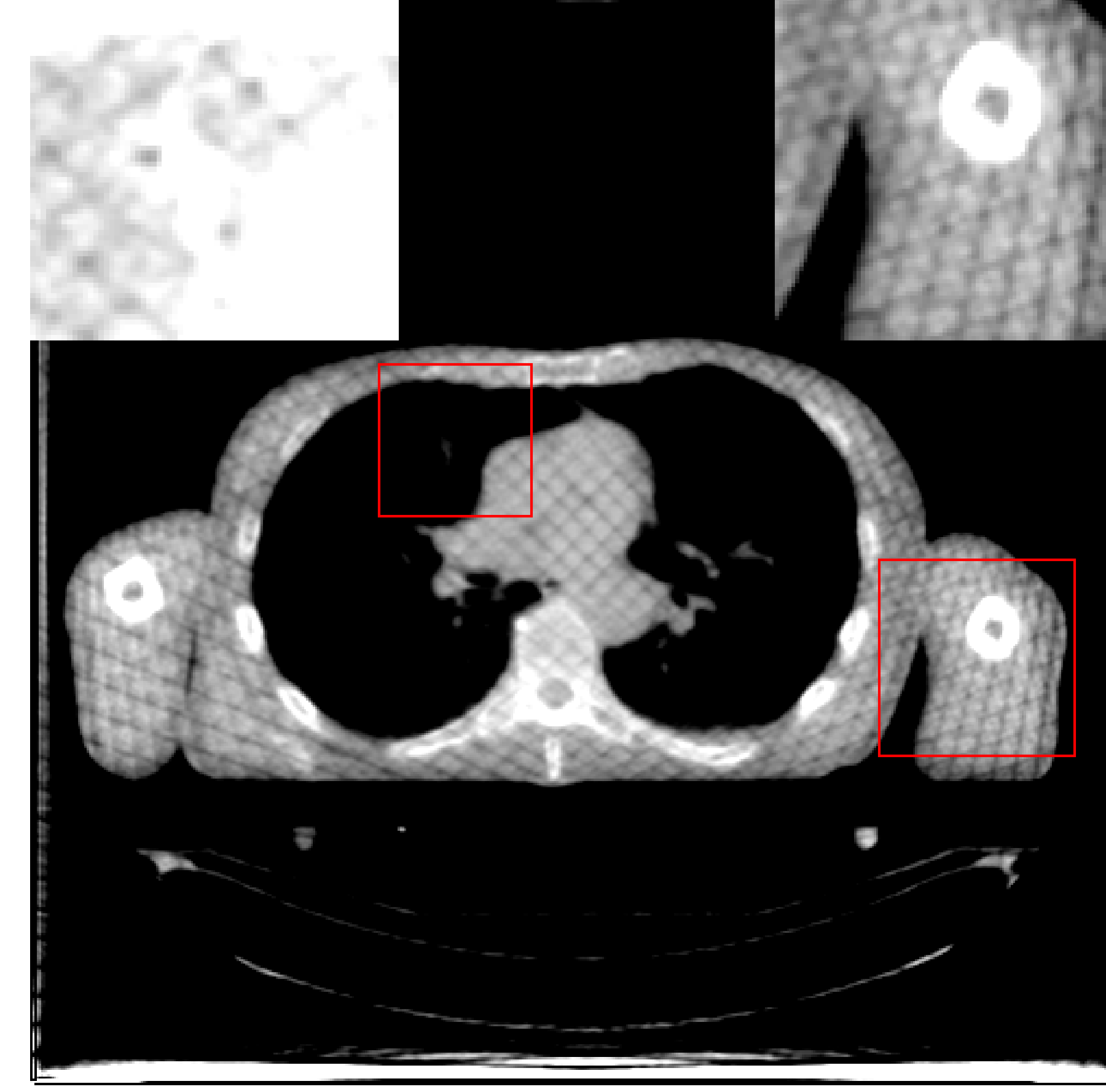}} \\
	\subfigure[lSIRT, 50 iterations]		{\includegraphics[width = 1.1in]{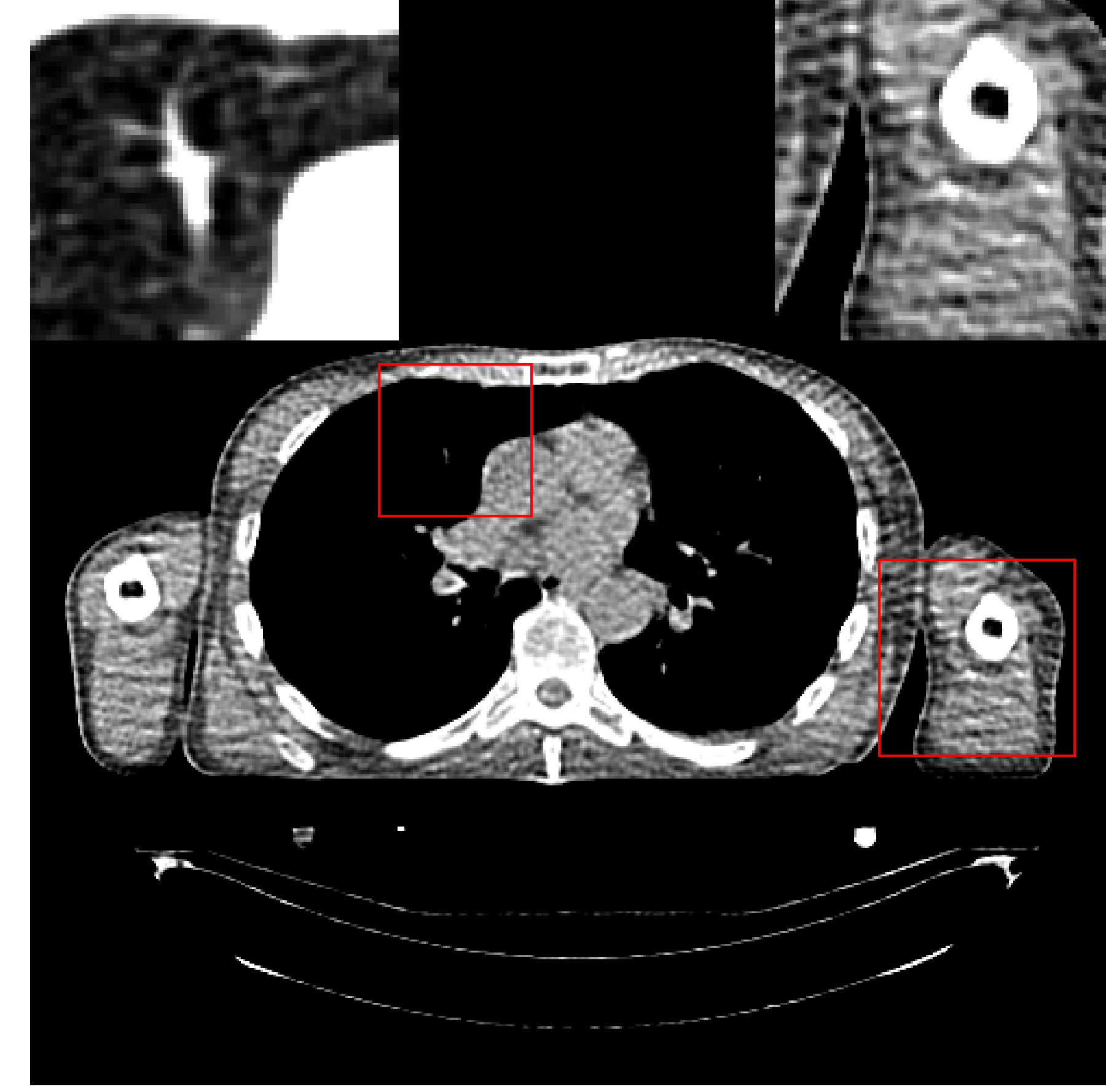}}
	\subfigure[lSIRT, 100 iterations]	{\includegraphics[width = 1.1in]{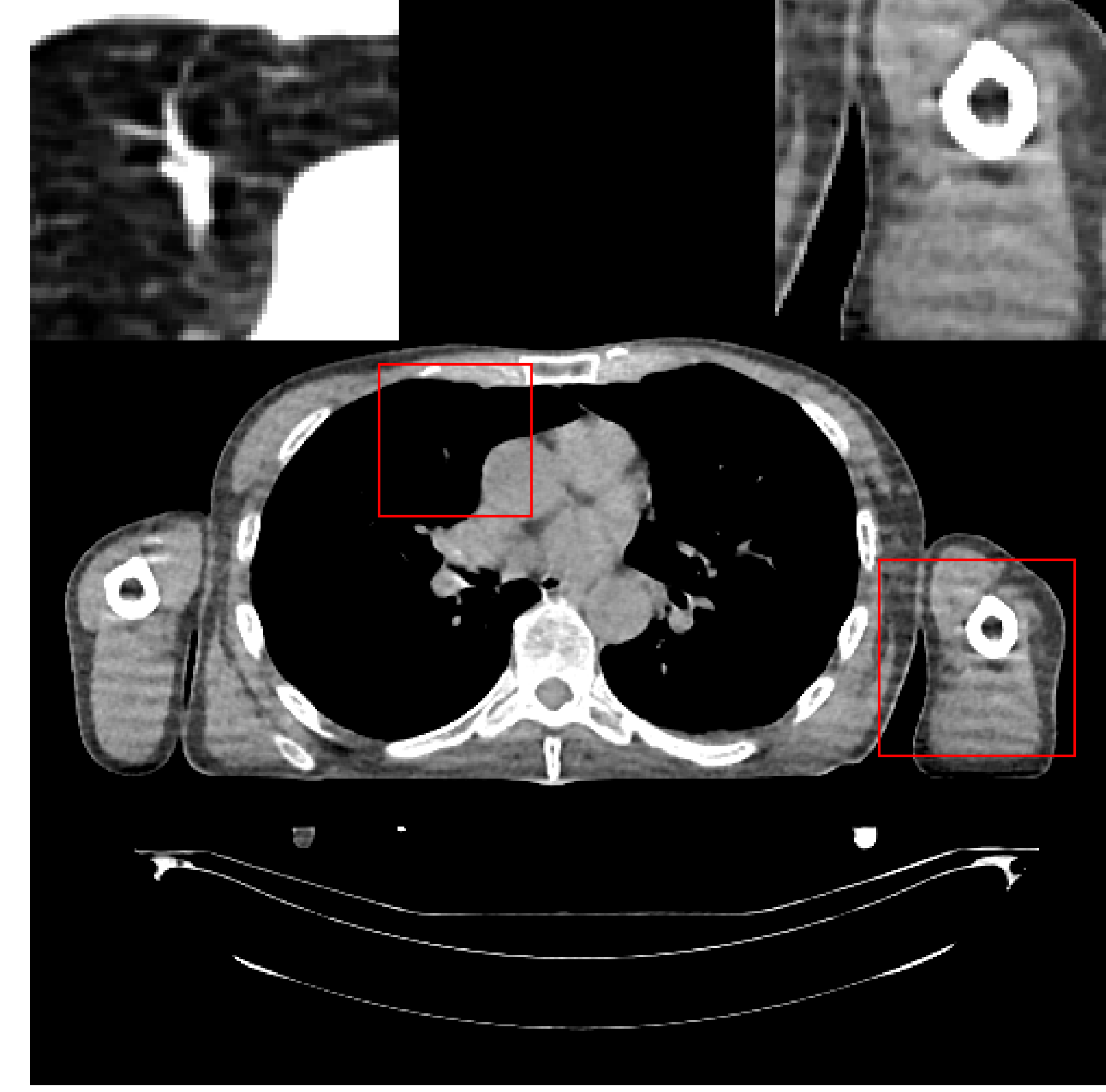}}
	\subfigure[lSIRT, 400 iterations]	{\includegraphics[width = 1.1in]{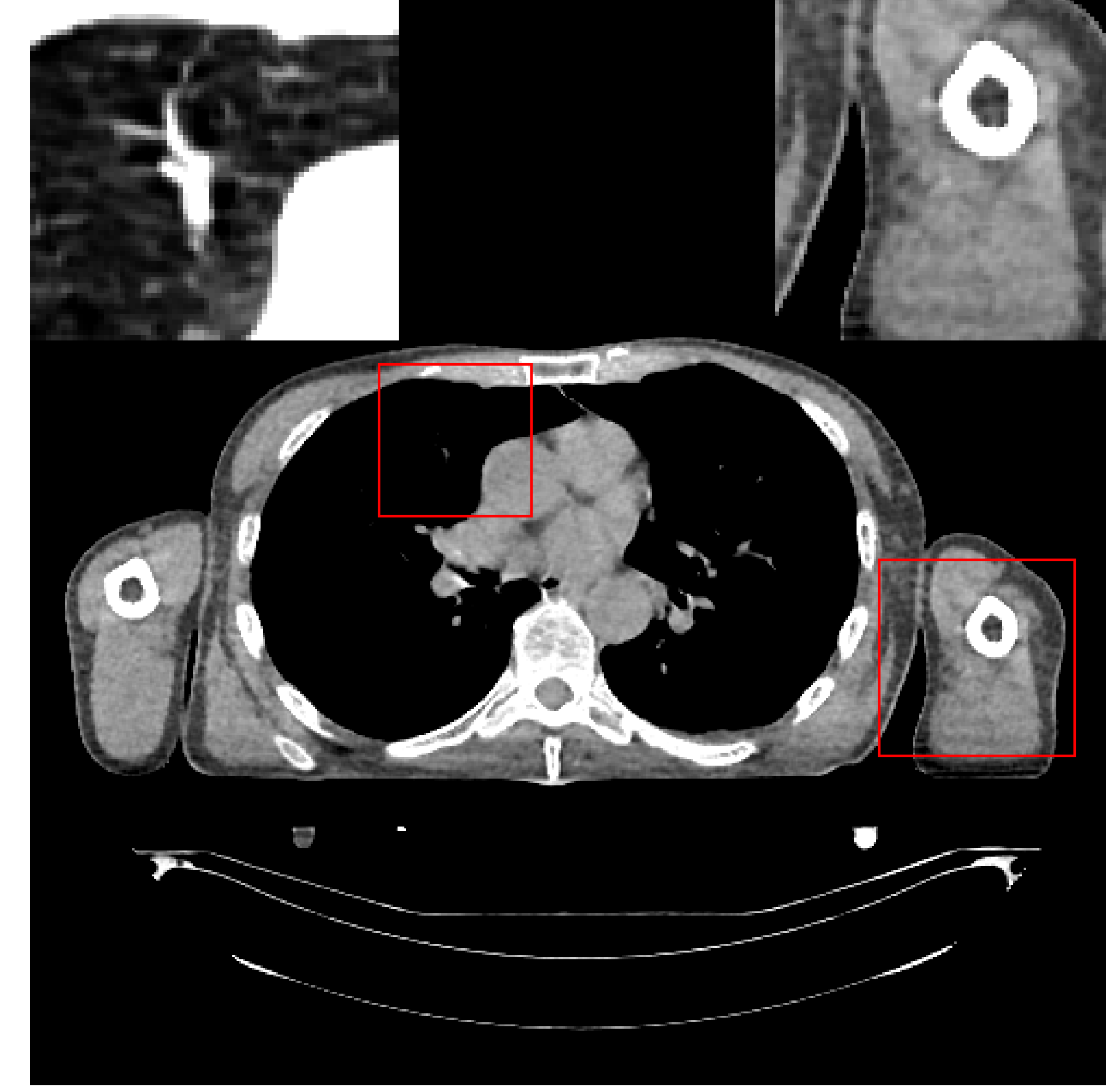}}
	\caption{The convergence of the 2D low-noise lSIRT model on a chest scan. For (a) and (d-f) A window of $-200 \, 200$HU was taken for the main image and right inset and a window of $-900 \, -400$HU was taken for the left inset everywhere. However, for (b-c), window levels were chosen to mimic the truth image.}
	\label{fig:lsirtconv}
\end{figure}

\begin{table}
\centering
	\caption{Reconstruction quality for models trained on 2D lung data}
\begin{tabular}{ l| l | c c c}
&Experiment & PSNR & SSIM &Runtime\\
&           & (dB) & & (ms)\\
	\hline
	\parbox[t]{2mm}{\multirow{7}{*}{\rotatebox[origin=c]{90}{Low Noise}}} 
	& 	FBP & $23.0$ & $0.585$ & $36$\\%
	&	SIRT & $28.5$ & $0.830$ & $720$\\%
	&	U-net & $36.1$ & $0.886$ & $731$\\%
	&	LPDorig & $\mathbf{42.8}$ & $\mathbf{0.964}$ & $106$\\%
	&	LPDsame & $42.0$ & $0.963$ & $111$\\%
	&	lSIRT* & $34.7$ & $0.927$ & $847$\\%
	&	lSIRT & $40.4$ & $0.957$ & $850$\\%
	\cline{1-5}
	
	\parbox[t]{2mm}{\multirow{7}{*}{\rotatebox[origin=c]{90}{High Noise}}} 
	& 	FBP & $13.4$ & $0.077$ & $36$\\%
	&	SIRT & $23.3$ & $0.355$ & $727$\\%
	&	U-net & $32.7$ & $0.862$ & $623$\\%
	&	LPDorig & $\mathbf{34.5}$ & $0.899$ & $107$\\%
	&	LPDsame & $33.6$ & $0.900$ & $108$\\%
	&	lSIRT* & $31.2$ & $0.841$ & $855$\\ %
	&	lSIRT & $31.7$ & $\mathbf{0.922}$ & $863$\\%
\hline
\end{tabular}
    \vspace{5mm}
    \newline
    Image quality statistics for the image reconstruction on patient data. The quantities are averaged over 100 different images in the 2D patient dataset.
\label{tab:patient}

\end{table}
In Figure~\ref{fig:lsirtconv} several examples of the reconstruction of lSIRT for a different number of iterations is given. Between 100 and 400 iterations certain artifacts are disappearing even though, the PSNR and SSIM do not improve much further. This is especially the case for the high-noise regime. Therefore we select 100 iterations as the trade-off between computation time and image quality throughout the rest of the article. For the same reason the choice of $N_s$ and $N_{\text{tot}}$, was made.

\subsubsection{3D models}

The models trained on the ellipse data were evaluated both on the ellipse data and the 3D Shepp Logan phantom. The performance is given in Table~\ref{tab:3D}. For CBCT reconstruction, the lSIRT algorithm performs clearly better than FBP and SIRT. We do not consider the lSIRT* algorithm, as the 2D results have shown that lSIRT is superior with a minimal computational overhead.

\begin{figure}
	\centering
	\begin{tabular}{cccc}
	&Shepp Logan & Head & Pelvis\\
\vspace{-3mm}
	\rotatebox[origin=c]{90}{\hspace{1cm}Truth} &
\hspace{-3mm}\subfigure                         {\includegraphics[width = 1.0in]{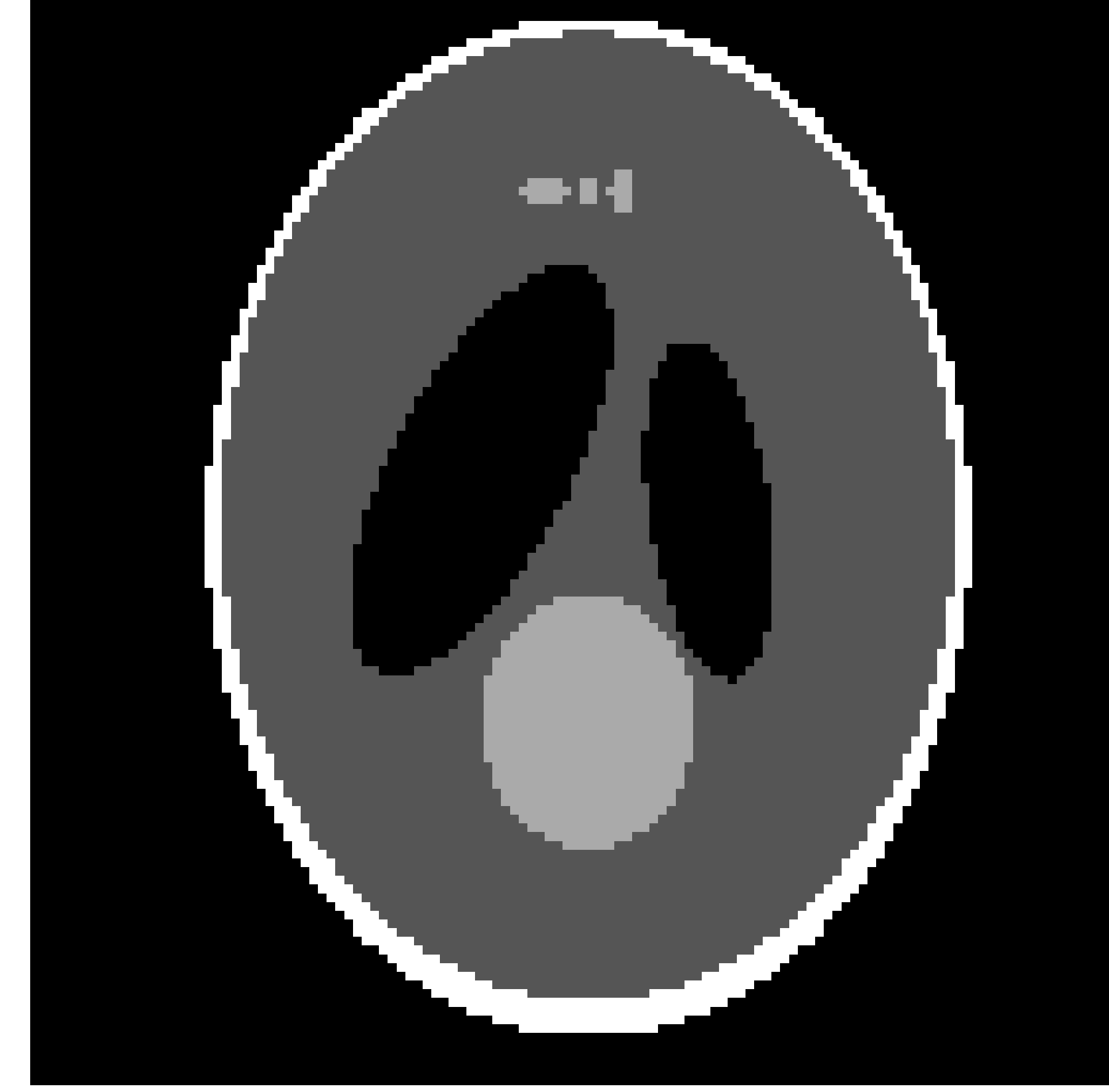}}&
\hspace{-3mm}\subfigure                         {\includegraphics[width = 1.0in]{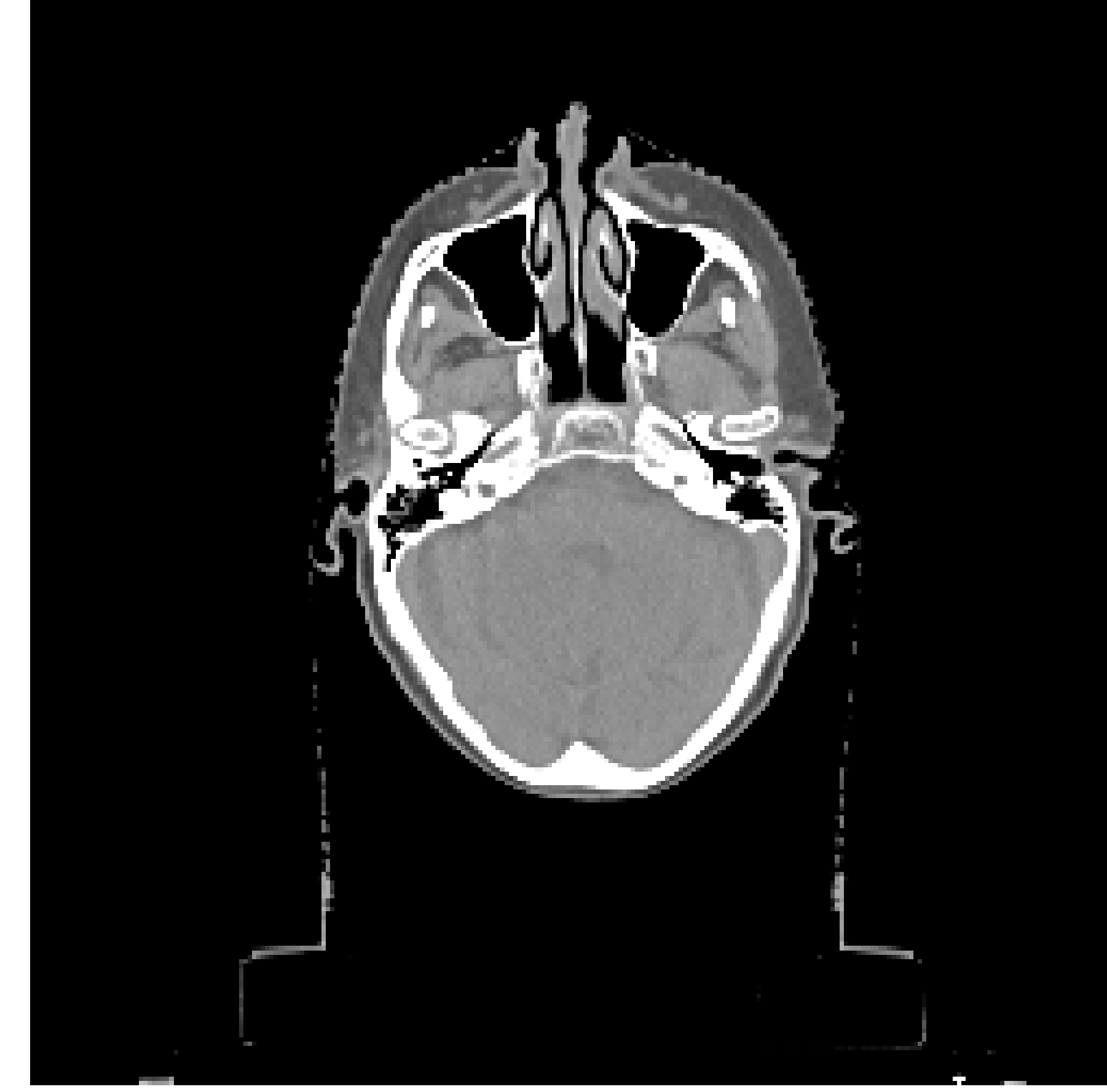}}&
\hspace{-3mm}\subfigure                         {\includegraphics[width = 1.0in]{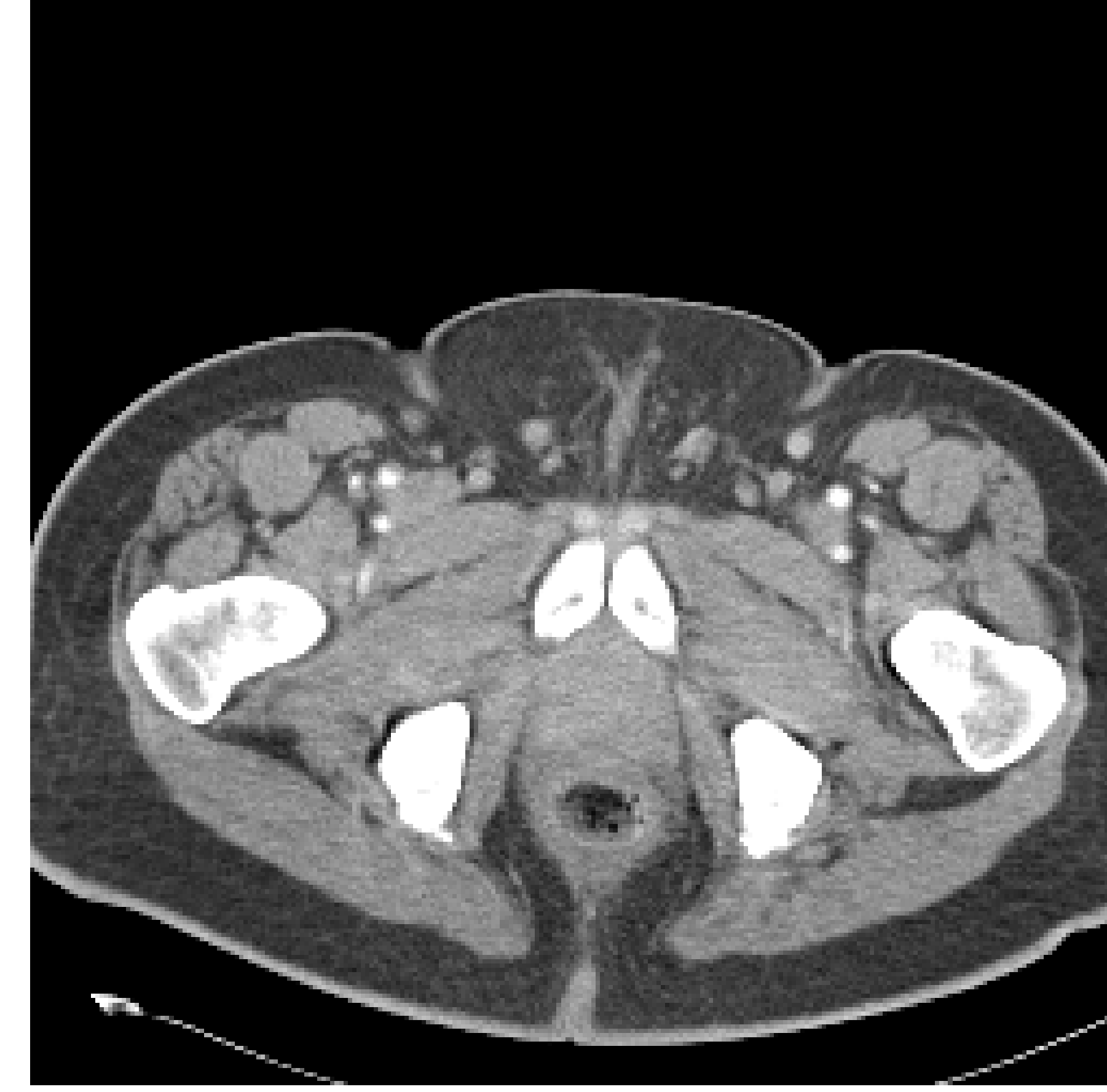}}\\
\vspace{-3mm}
	\rotatebox[origin=c]{90}{\hspace{1cm}FBP} &
\hspace{-3mm}\subfigure                         {\includegraphics[width = 1.0in]{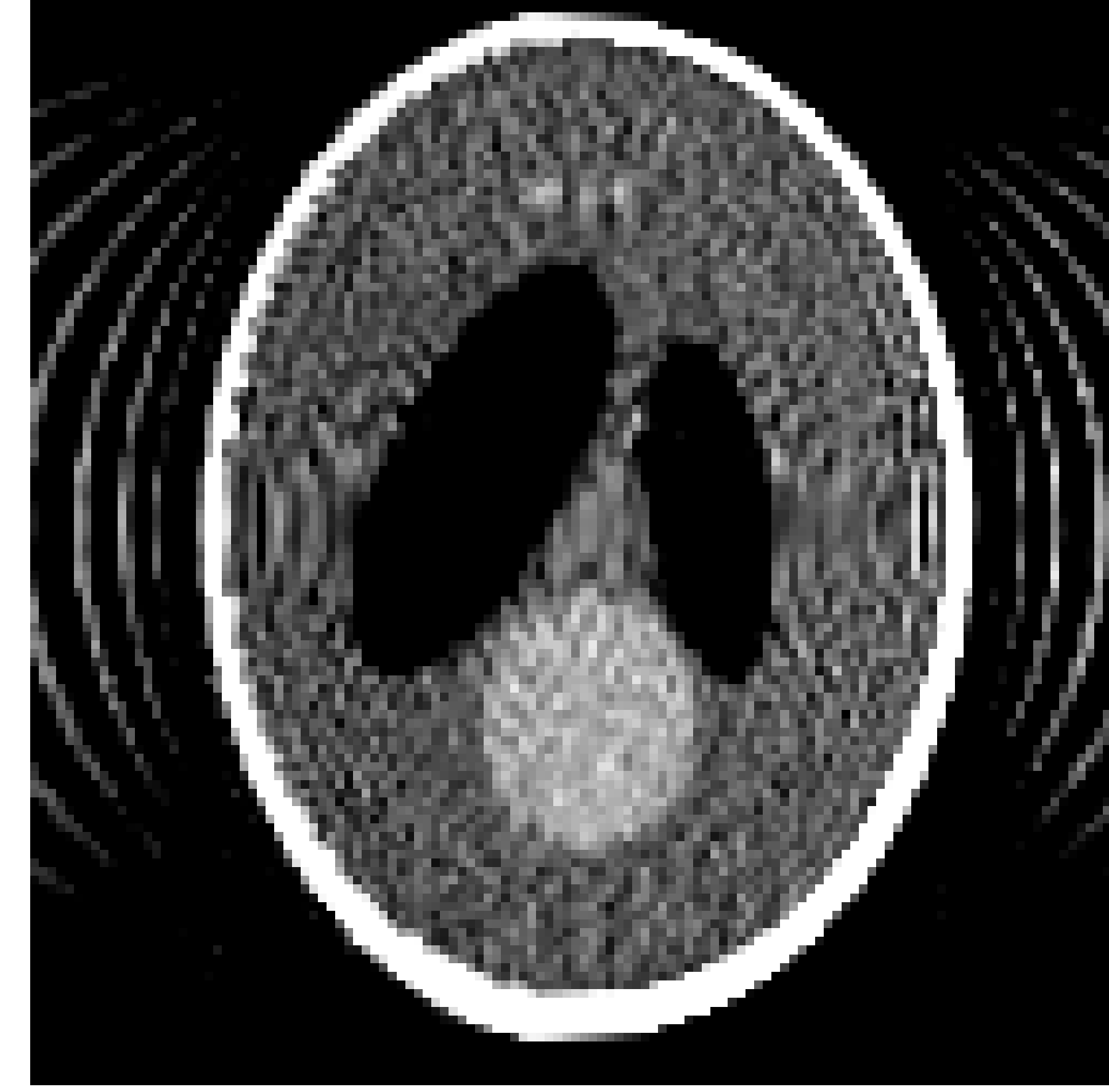}}&
\hspace{-3mm}\subfigure                         {\includegraphics[width = 1.0in]{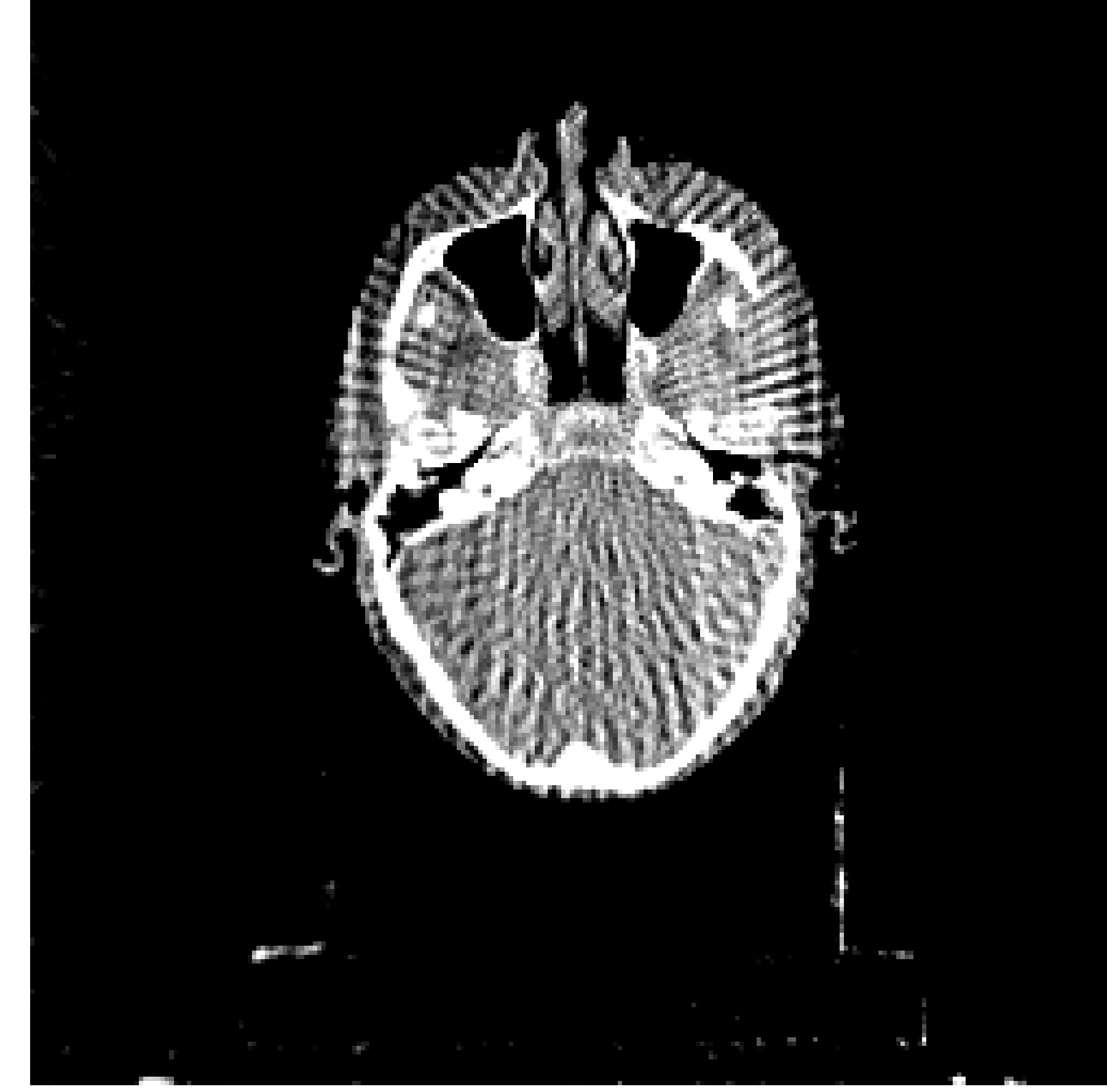}}&
\hspace{-3mm}\subfigure                         {\includegraphics[width = 1.0in]{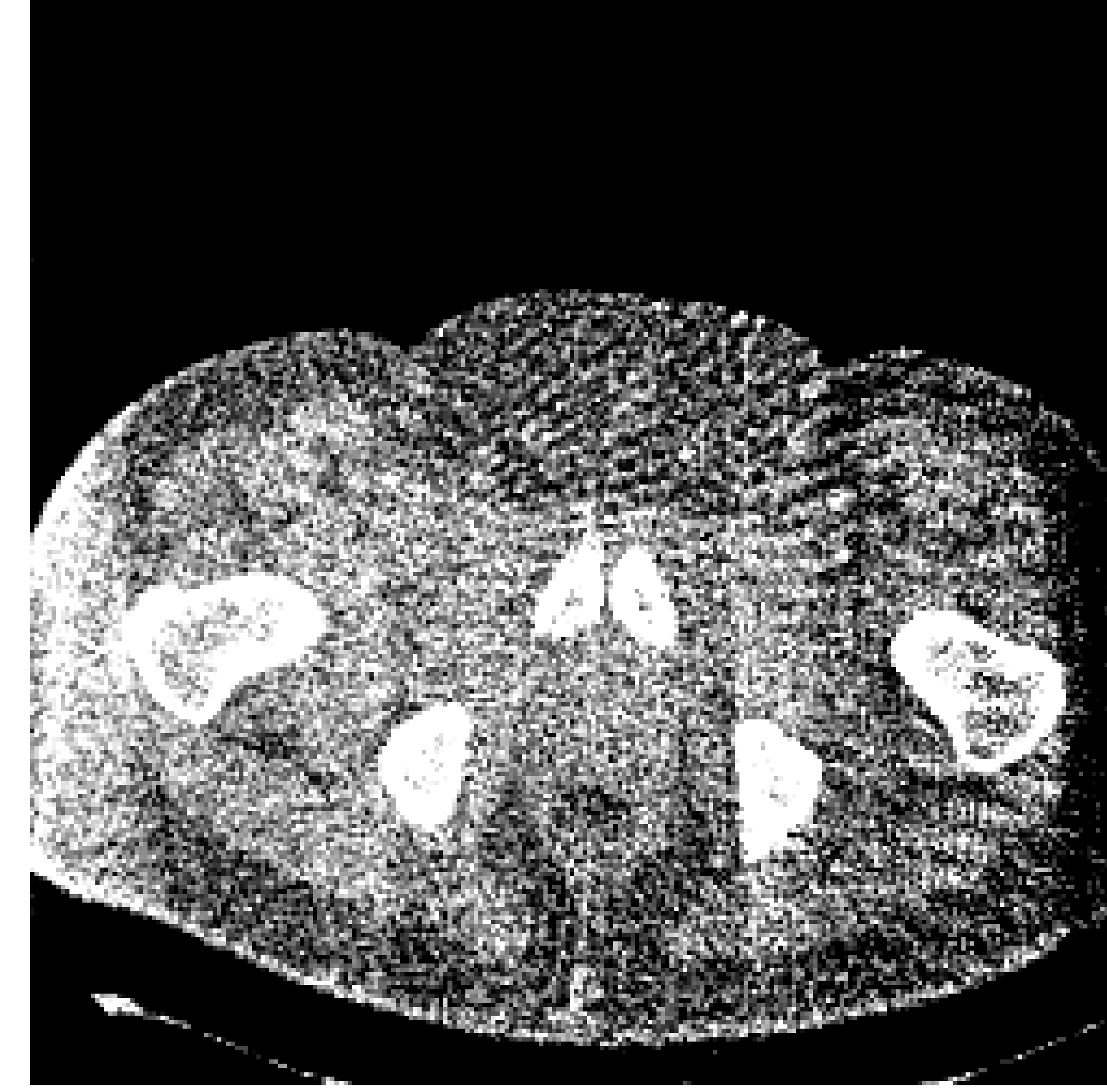}}\\
\vspace{-3mm}
	\rotatebox[origin=c]{90}{\hspace{1cm}SIRT} &
\hspace{-3mm}\subfigure                         {\includegraphics[width = 1.0in]{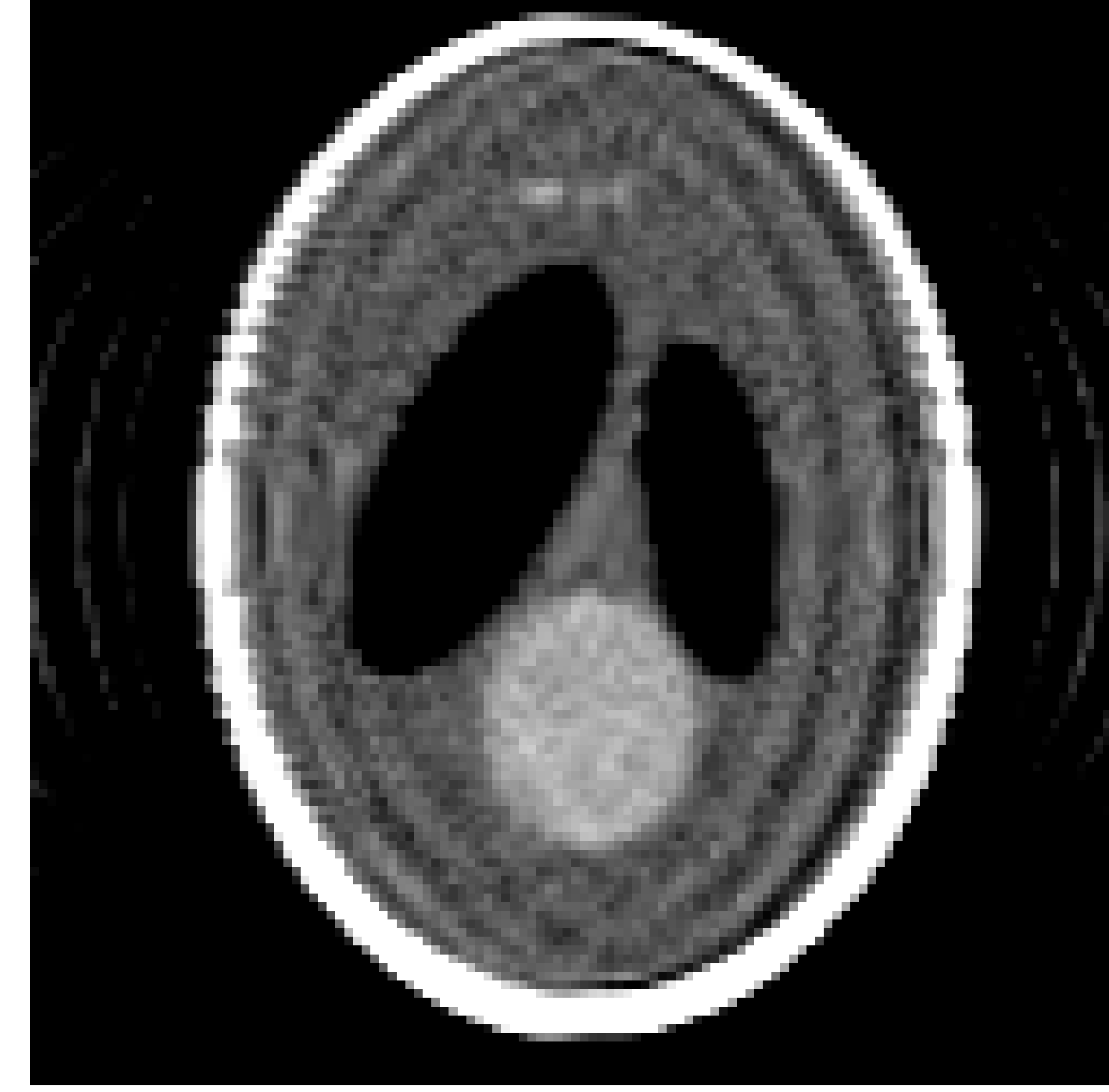}}&
\hspace{-3mm}\subfigure                         {\includegraphics[width = 1.0in]{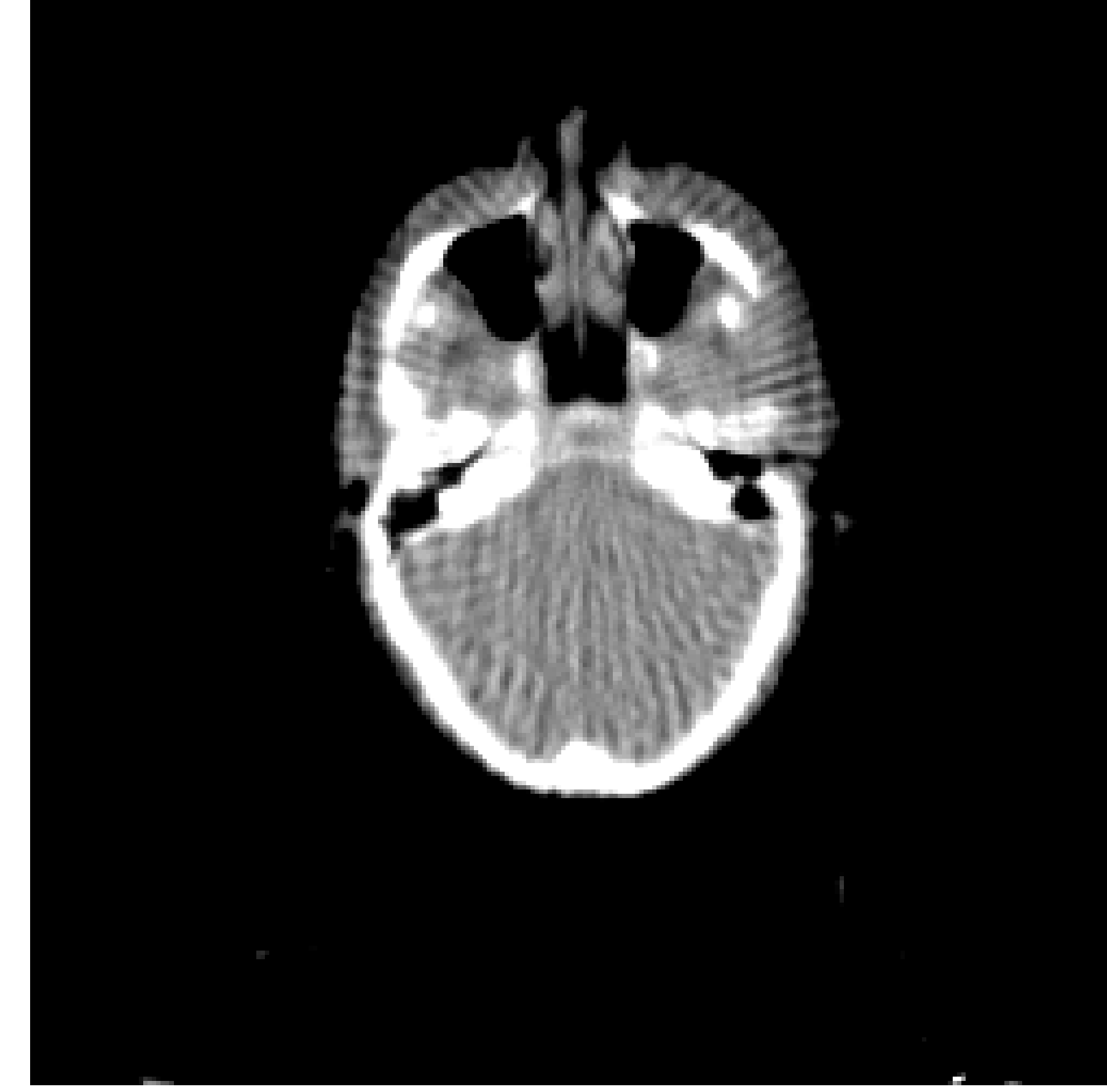}}&
\hspace{-3mm}\subfigure                         {\includegraphics[width = 1.0in]{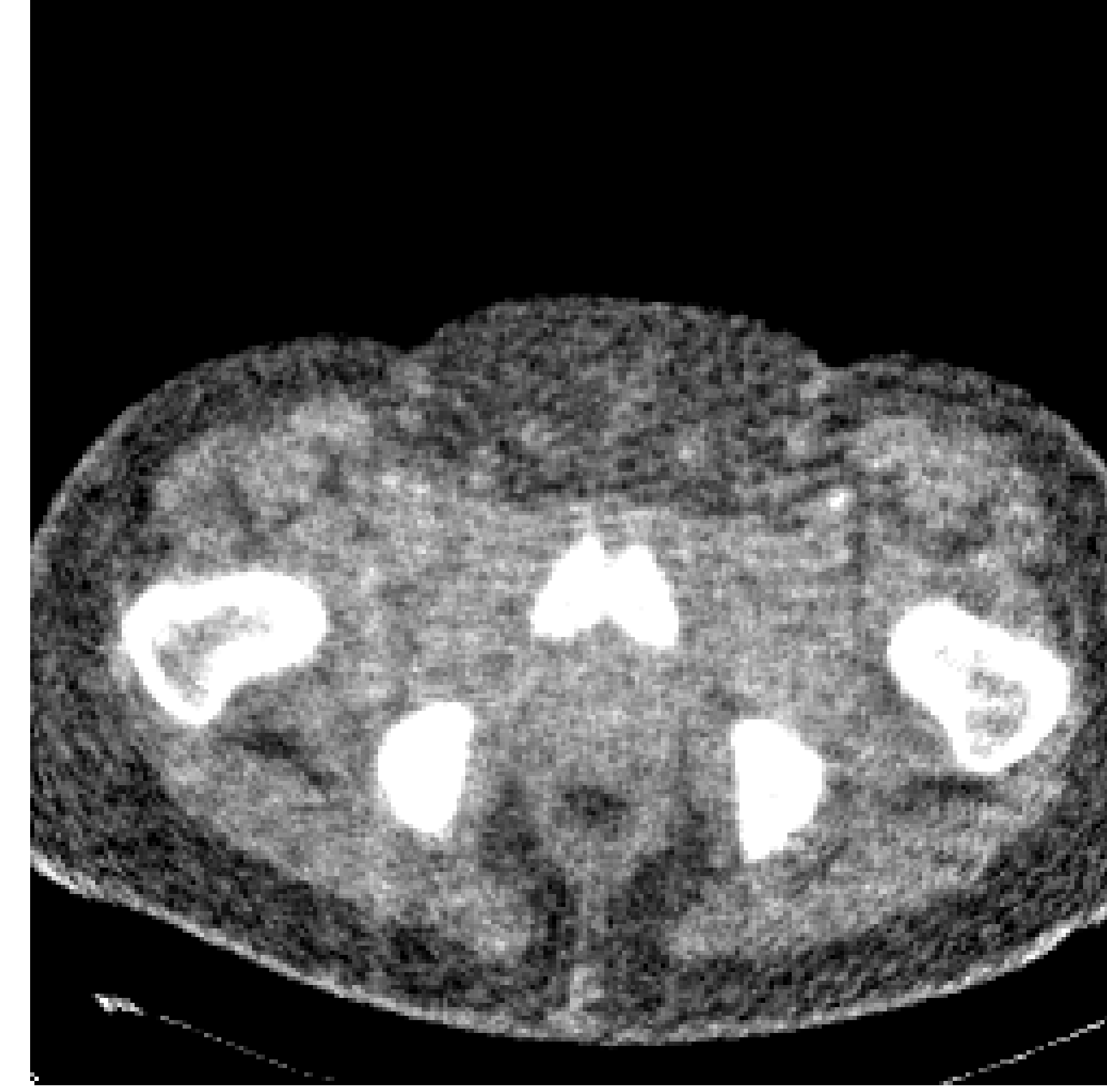}}\\
	\rotatebox[origin=c]{90}{\hspace{1cm}lSIRT} &
\hspace{-3mm}\subfigure                         {\includegraphics[width = 1.0in]{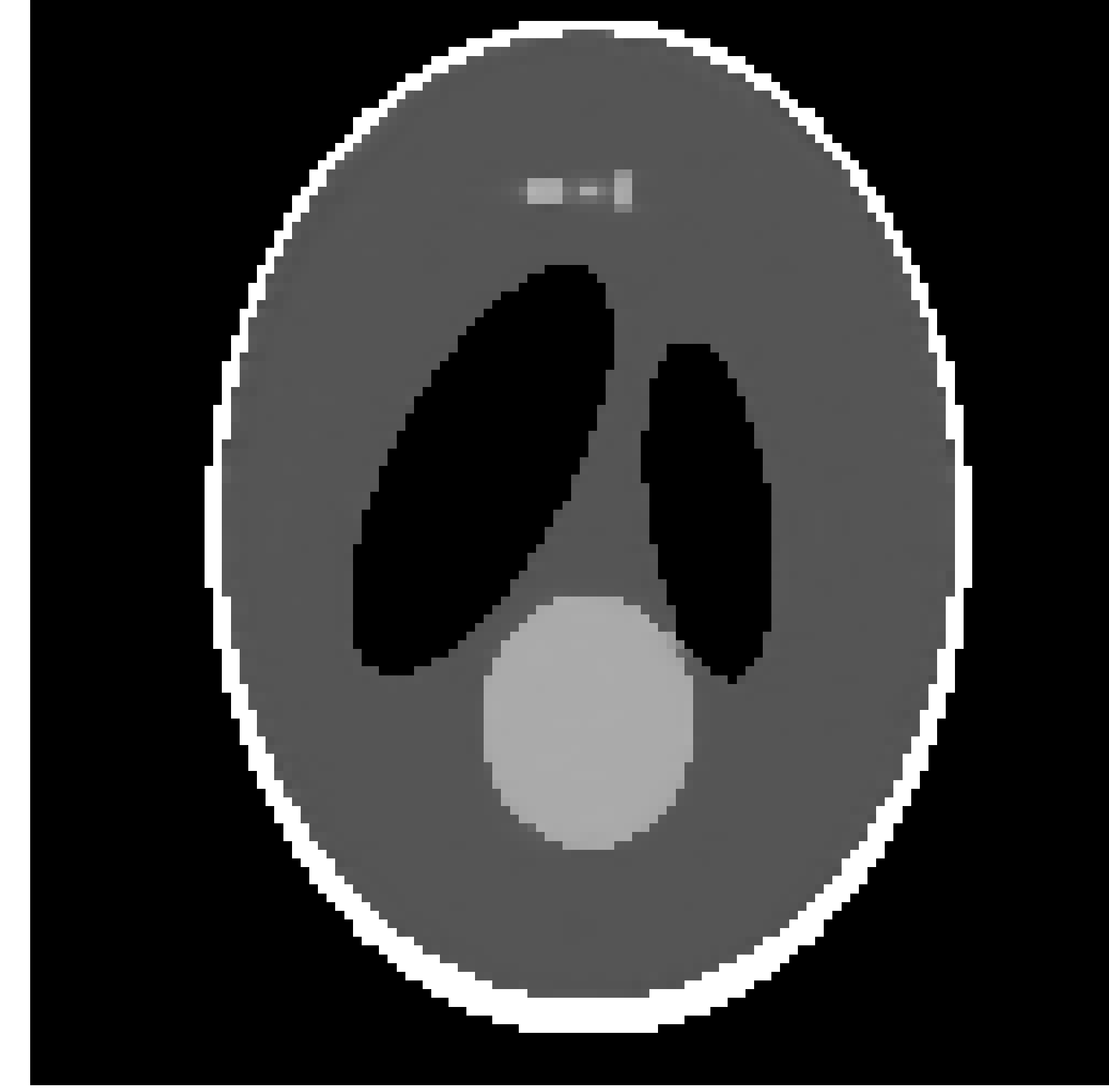}}&
\hspace{-3mm}\subfigure                         {\includegraphics[width = 1.0in]{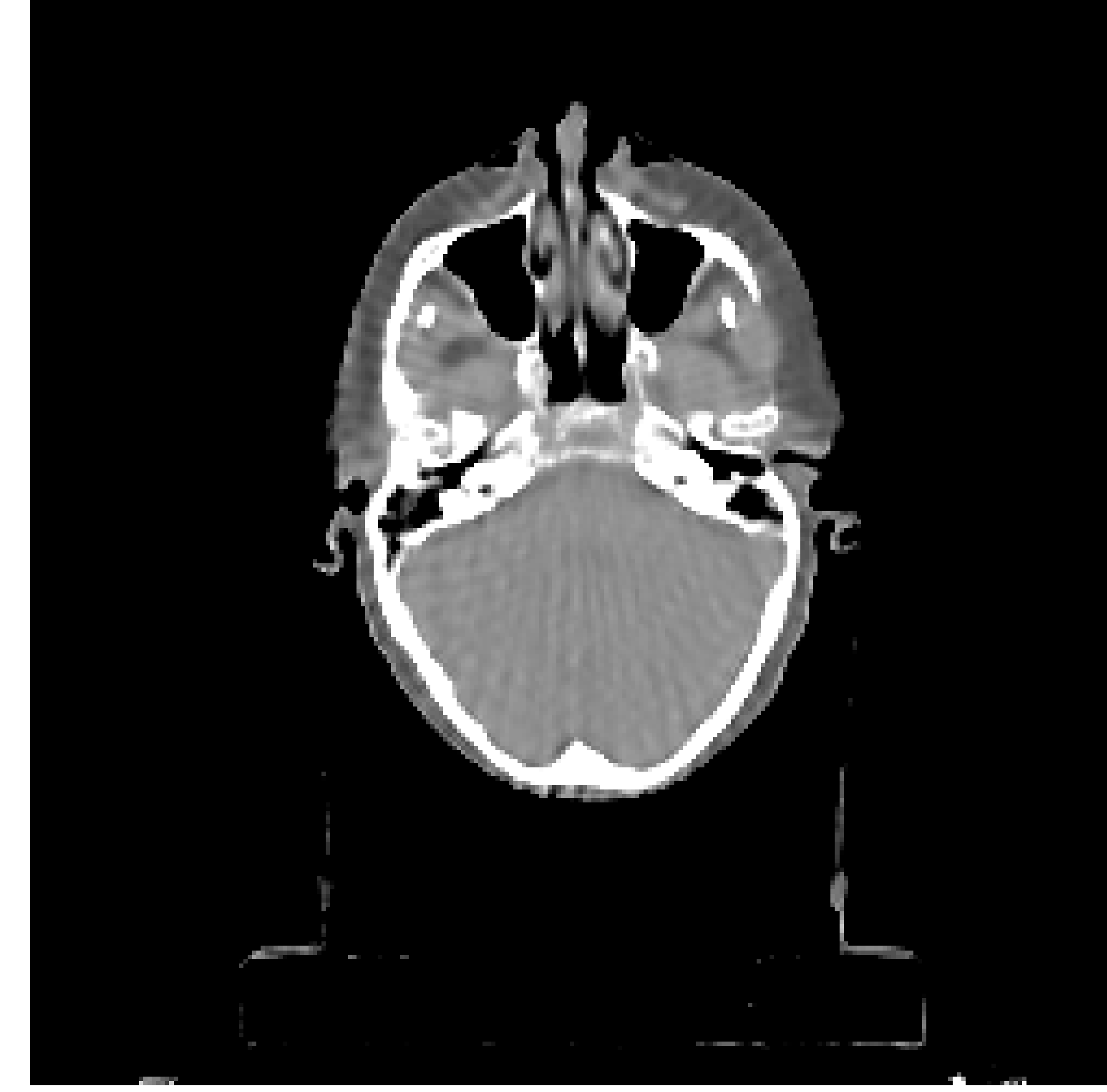}}&
\hspace{-3mm}\subfigure               	        {\includegraphics[width = 1.0in]{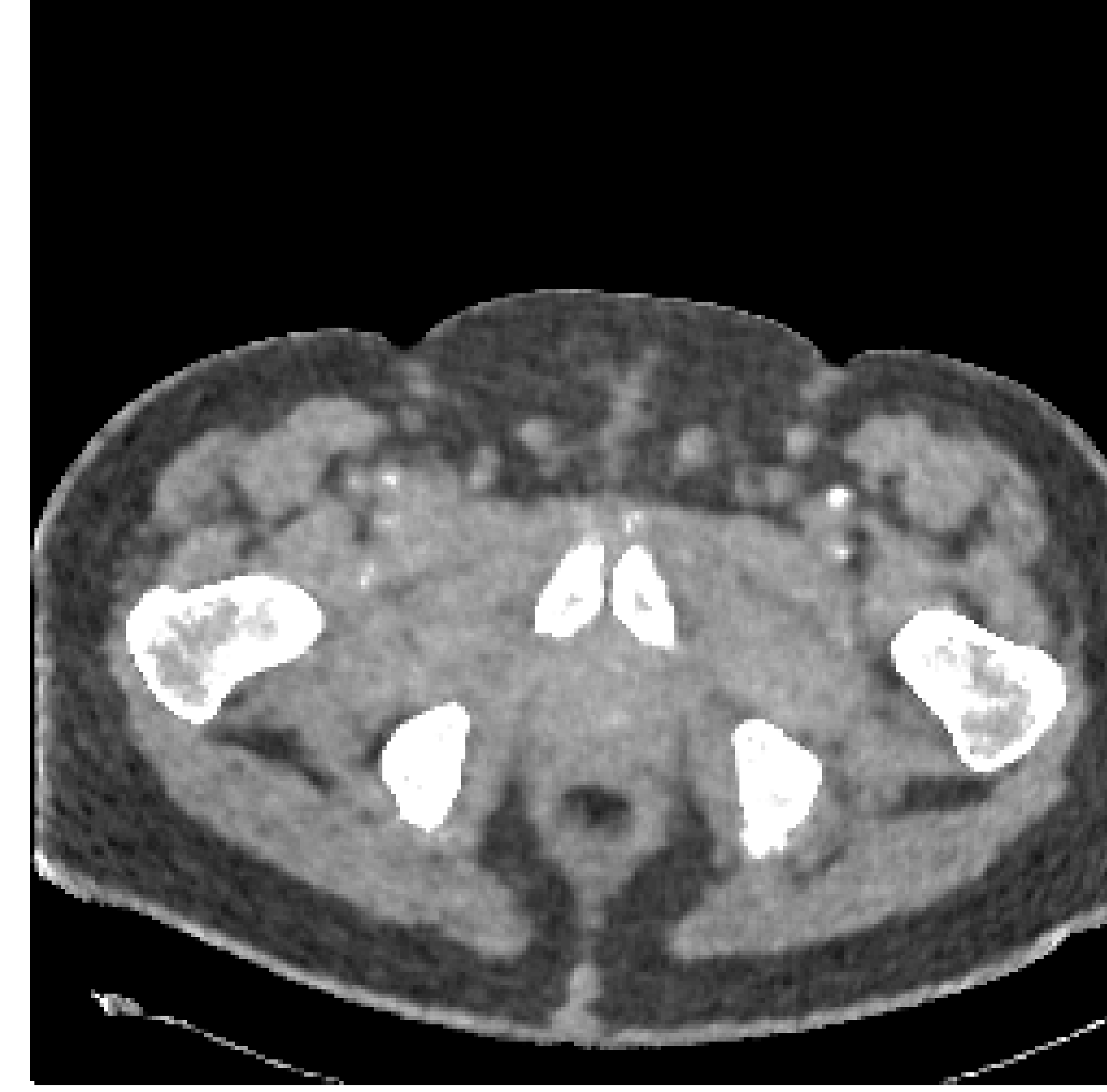}}
\end{tabular}

	\caption{CBCT Reconstructions. In the first column: the 3D Shepp Logan phantom on $128^3$ voxels under low-noise conditions (window $[100, 400] \text{HU}$). The second column contains, a low-noise head reconstruction with $256^3$ voxels from artificial projection data, window $[-300, 300] \text{HU}$. The final third column displays, a low-noise pelvis reconstruction with $256^3$ voxels from artificial projection data (window $[-1200, 300]\text{HU}$). }
	\label{fig:cbct}
\end{figure}

Figure~\ref{fig:cbct} provides examples for the low noise regime for FBP, SIRT and lSIRT when trained on the lung data. Examples are given for the 3D Shepp Logan Phantom, a CT scan including (part of) the head, and a scan of the pelvis. Of note is that the head is a scan taken from the testing set of the lung data, while the pelvis data was from an independent dataset \cite{TCIA,Zuley16} of CT scans of the pelvis, which were not seen during training. The corresponding performance metrics are given in Table~\ref{tab:3D}, showing a superior performance compared to the classic methods. Compared to the SIRT and FBP, lSIRT is the superior image reconstruction method. We obtain PSNRs of $26.0\text{dB}$, $31.1\text{dB}$ and SSIMs of $0.702$, $0.915$ for SIRT and lSIRT respectively.

\begin{figure}
	\centering
	\subfigure[truth]                         {\includegraphics[width = 1.1in]{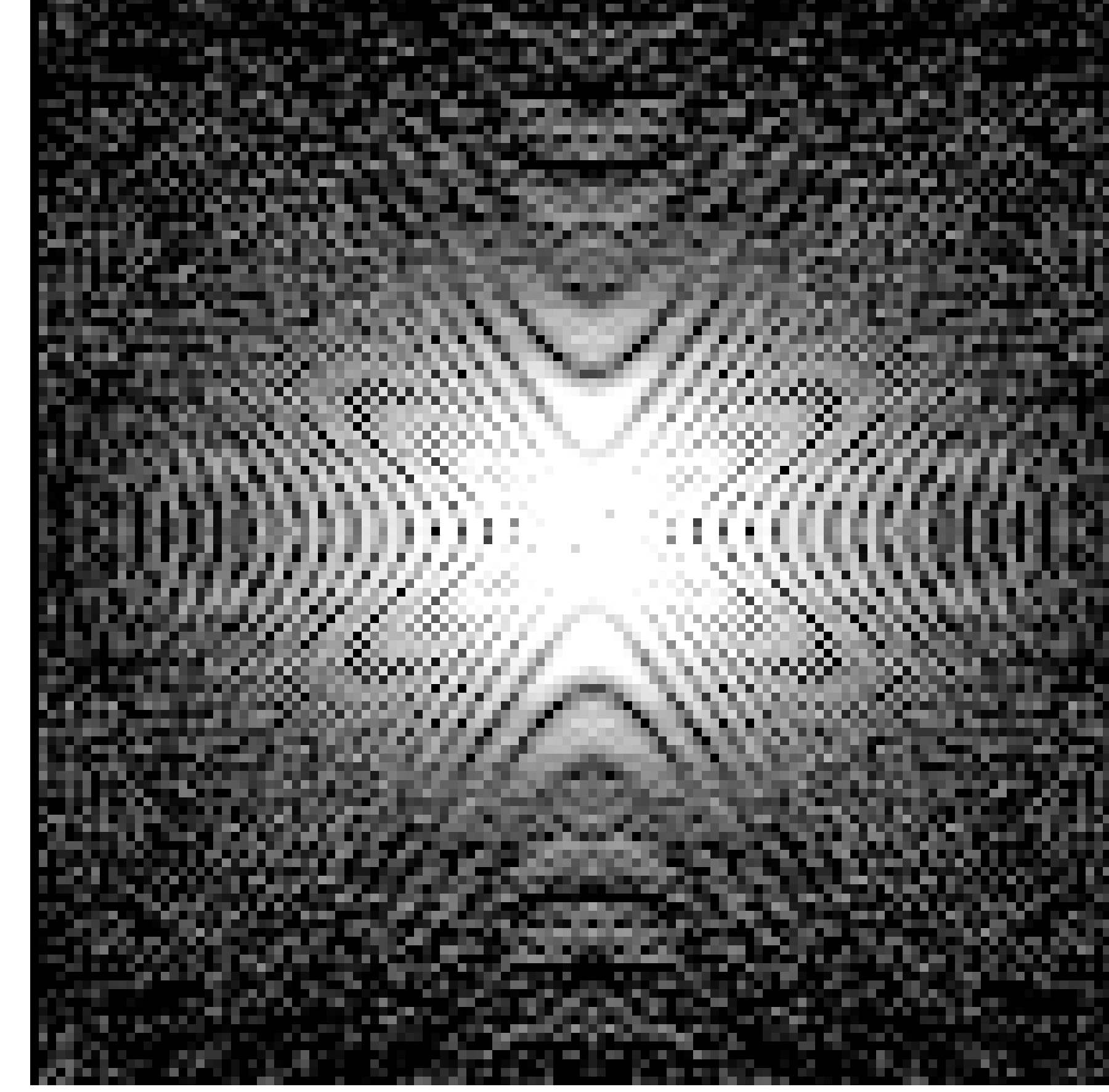}}
	\subfigure[FBP]                           {\includegraphics[width = 1.1in]{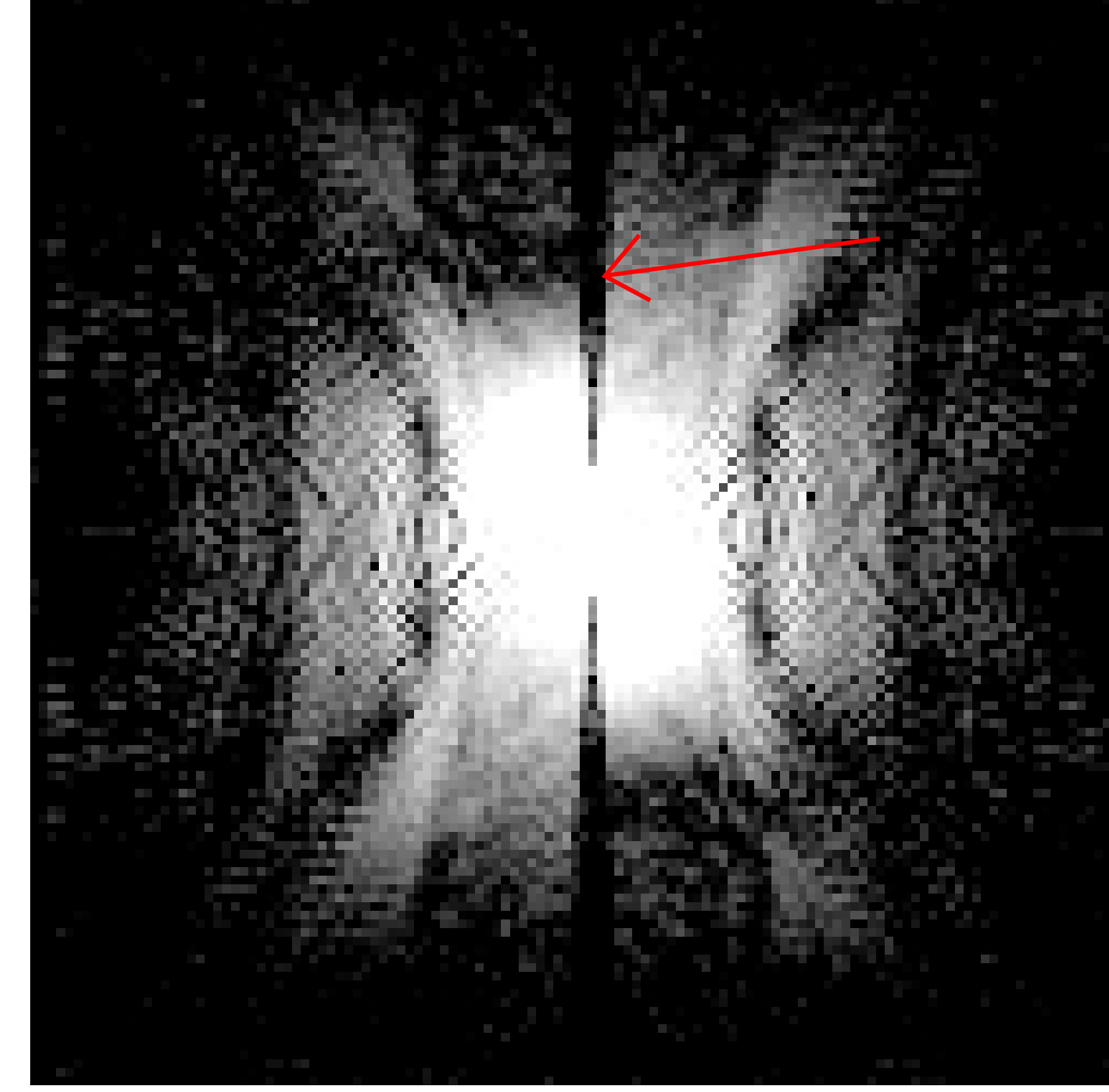}}\\
	\subfigure[SIRT]                          {\includegraphics[width = 1.1in]{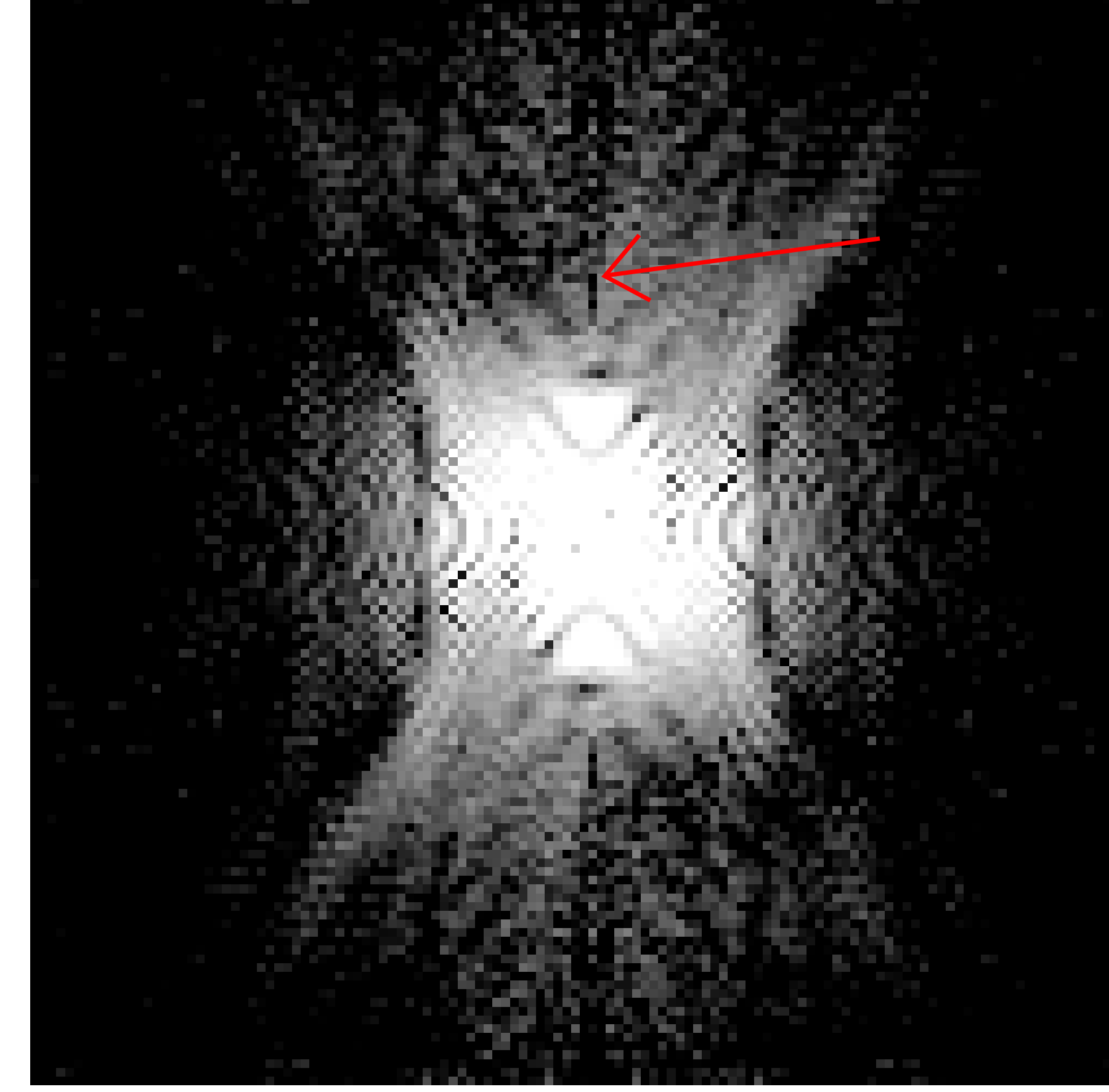}}
 	\subfigure[learned SIRT]                  {\includegraphics[width = 1.1in]{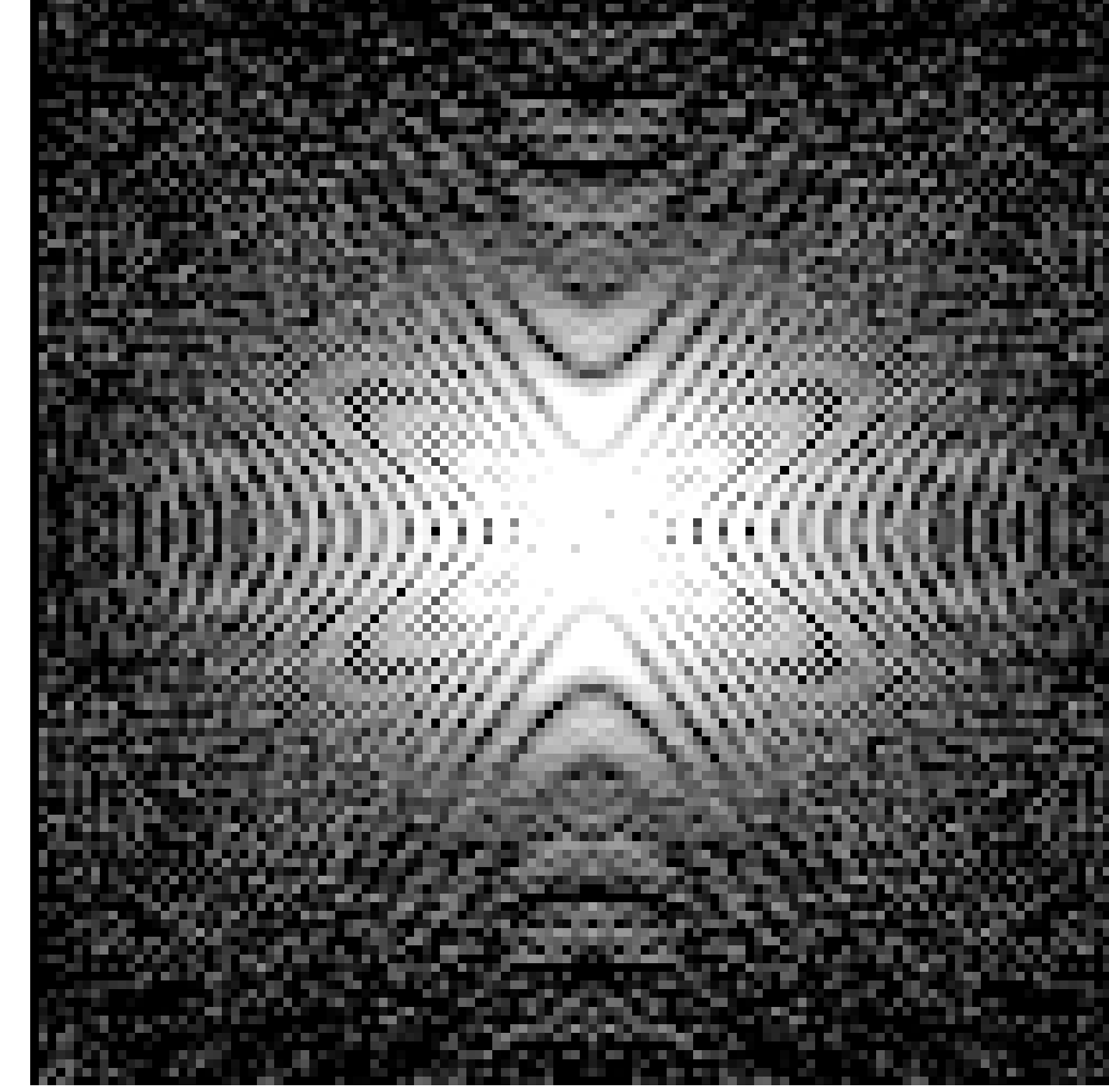}}\\

	\caption{The discrete Fourier transform of the 3D Shepp Logan reconstructions in Figure~\ref{fig:cbct}. Notice the stripes in the FBP and SIRT reconstructions, that constitute the DeFrise artifact.}
	\label{fig:fft}
\end{figure}

The Defrise artifact results from the fact that parts of the image are illuminated only under an angle, which yields a lower resolution in the direction of the axis of rotation.
It is most visible in the Fourier domain, where two cones are not sampled.
In Figure~\ref{fig:fft} we show the discrete Fourier transform of the reconstructions of the Shepp Logan phantom.
For the FBP reconstruction, the Defrise artifact is very prominent as a black streak through in the Fourier domain of a coronal slice. For the SIRT reconstruction the artifact is also visible, but in the lSIRT reconstruction this cone appears to have disappeared.

\begin{table}
	\caption{Reconstruction quality for 3D models}
\begin{tabular}{l| l| l | c c c}
&{\rotatebox[origin=c]{90}{Noise}}&Experiment & PSNR & SSIM & Runtime\\
&&           & (dB) &      &  (s)\\
	\hline
\parbox[t]{2mm}{\multirow{9}{*}{\rotatebox[origin=c]{90}{Ellipse phantoms}}}&
	\parbox[t]{2mm}{\multirow{3}{*}{\rotatebox[origin=c]{90}{Low}}} 
	 &	FBP & 25.1 & 0.441 & 0.86\\
	&&	SIRT & 28.7 & 0.601 &  6.4\\
	&&	lSIRT & $\mathbf{51.2}$ & $\mathbf{0.998}$ & 10.4\\
	\cline{2-6}
	&
	\parbox[t]{2mm}{\multirow{3}{*}{\rotatebox[origin=c]{90}{Medium}}} 
	 &	FBP & 24.4 & 0.382 &  0.83 \\
	&&	SIRT & 27.5 & 0.595 &  6.6 \\
	&&	lSIRT & $\mathbf{42.4}$& $\mathbf{0.987}$ &  10.4 \\
	\cline{2-6}
	&
	\parbox[t]{2mm}{\multirow{3}{*}{\rotatebox[origin=c]{90}{High}}} 
	 &	FBP & 22.9 & 0.323   &0.52\\
	&&	SIRT & 24.3& 0.490  & 5.6\\
	&&	lSIRT & $\mathbf{37.0}$& $\mathbf{0.945}$ &9.4\\
\hline
\parbox[t]{2mm}{\multirow{9}{*}{\rotatebox[origin=c]{90}{3D Shepp Logan}}}&
	\parbox[t]{2mm}{\multirow{3}{*}{\rotatebox[origin=c]{90}{Low}}} 
	 &	FBP & 20.1 & 0.489  & 0.91\\
	&&	SIRT & 21.6 & 0.718  & 6.4\\
	&&	lSIRT & $\mathbf{48.2}$&$\mathbf{ 0.999}$  & 10.7\\
	\cline{2-6}
	&
	\parbox[t]{2mm}{\multirow{3}{*}{\rotatebox[origin=c]{90}{Medium}}} 
	 &	FBP & 18.4 & 0.364  & 0.79 \\
	&&	SIRT & 18.5 & 0.581  & 6.3 \\
	&&	lSIRT & $\mathbf{29.7}$& $\mathbf{0.973}$ & 10.7 \\
	\cline{2-6}
	&
	\parbox[t]{2mm}{\multirow{3}{*}{\rotatebox[origin=c]{90}{High}}} 
	 &	FBP & 15.6 & 0.332 &  0.83\\
	&&	SIRT & 16.8& 0.520 &  6.4\\
	&&	lSIRT & $\mathbf{25.5}$& $\mathbf{0.927}$& 10.4\\
\hline
\parbox[t]{2mm}{\multirow{3}{*}{\rotatebox[origin=c]{90}{Patient}}}&
	\parbox[t]{2mm}{\multirow{3}{*}{\rotatebox[origin=c]{90}{Low}}} 
	 &	FBP & 22.4 & 0.433 &  3.9 \\
	&&	SIRT & 30.9 & 0.744 &  41 \\
	&&	lSIRT & $\mathbf{36.5}$& $\mathbf{0.943}$ &   90  \\
\hline

\end{tabular}
    \vspace{5mm}
    \\
	For the ellipses, these values pertain to an average over 10 different phantoms. For the Shepp Logan phantom, the image is kept the same, but an average is taken over 10 noise realizations in the sinogram. 
\label{tab:3D}
\end{table}

%

%

%

%

\section{Phantom measurements on CBCT scanner}
For comparison with the several learned approaches, FBP with and without a Hann filter window and SIRT with different number of iterations are reconstructed.

\begin{figure}
	\centering
	\subfigure[FPB no filter]          	{\includegraphics[width = 1.1in]{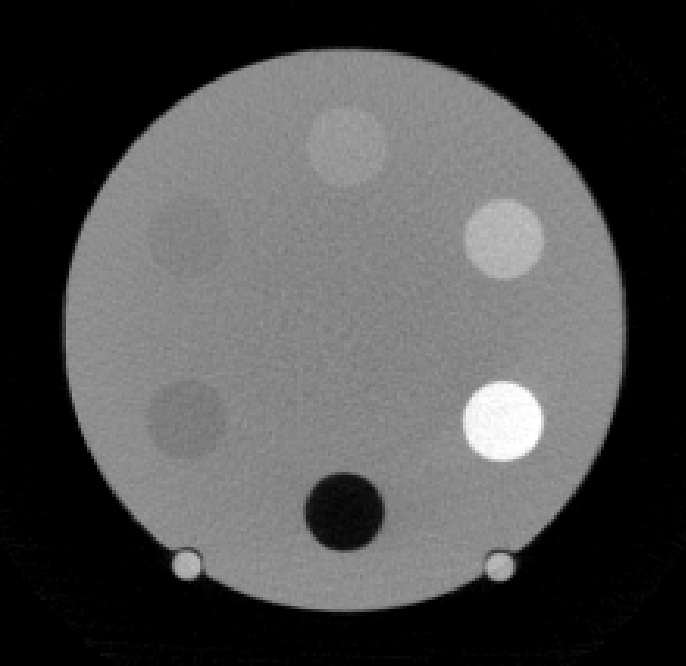}}
	\subfigure[FBP h=0.8]   		    {\includegraphics[width = 1.1in]{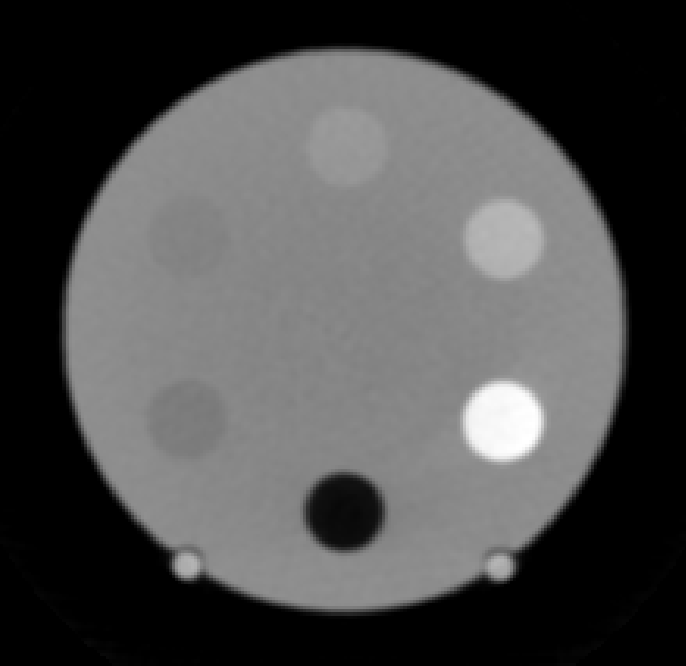}}
	\subfigure[SIRT 250 iterations]		{\includegraphics[width = 1.1in]{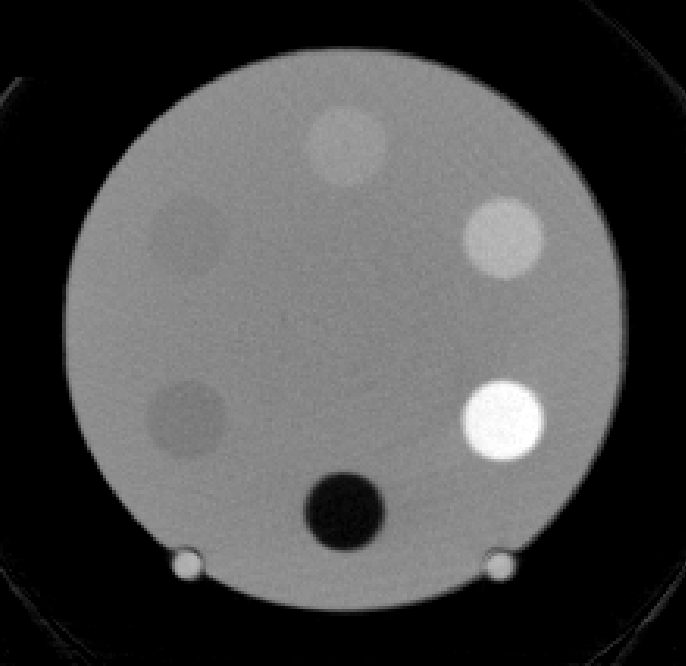}}\\
	\subfigure[lSIRT, low noise model]	{\includegraphics[width = 1.1in]{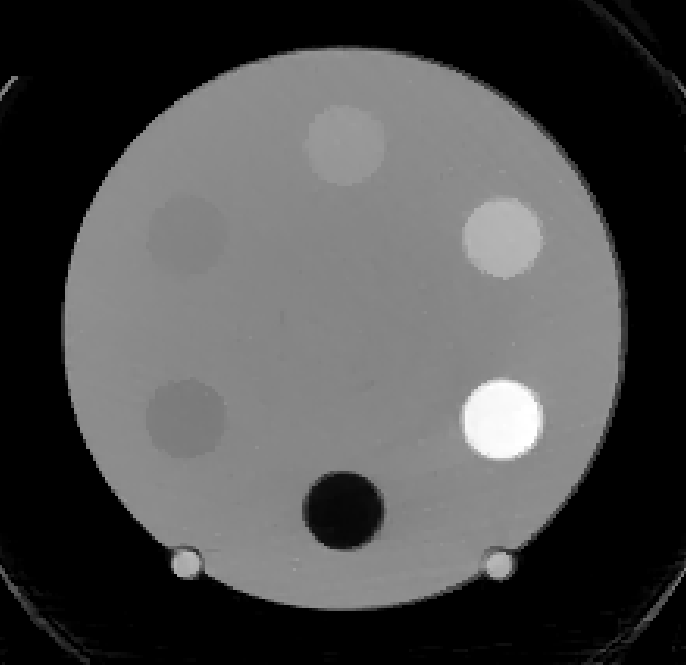}}
	\subfigure[lSIRT, highnoise model]	{\includegraphics[width = 1.1in]{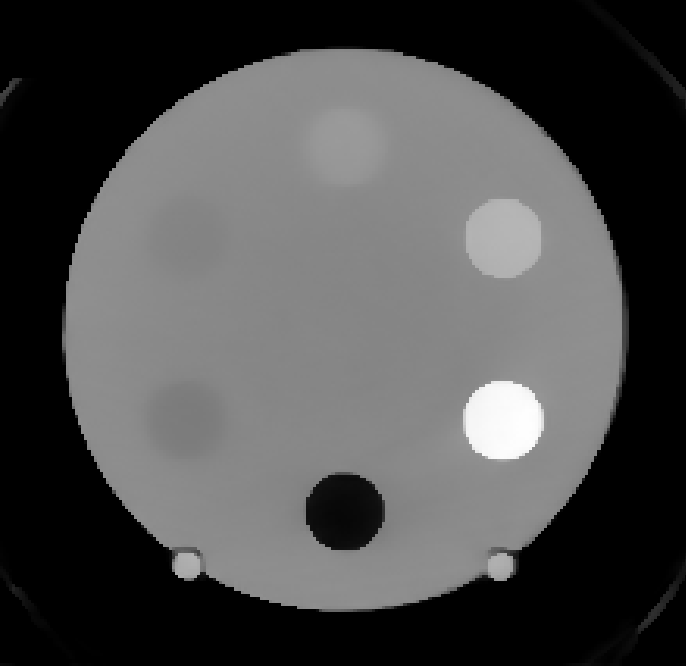}}
	\subfigure[lSIRT, patient model]	{\includegraphics[width = 1.1in]{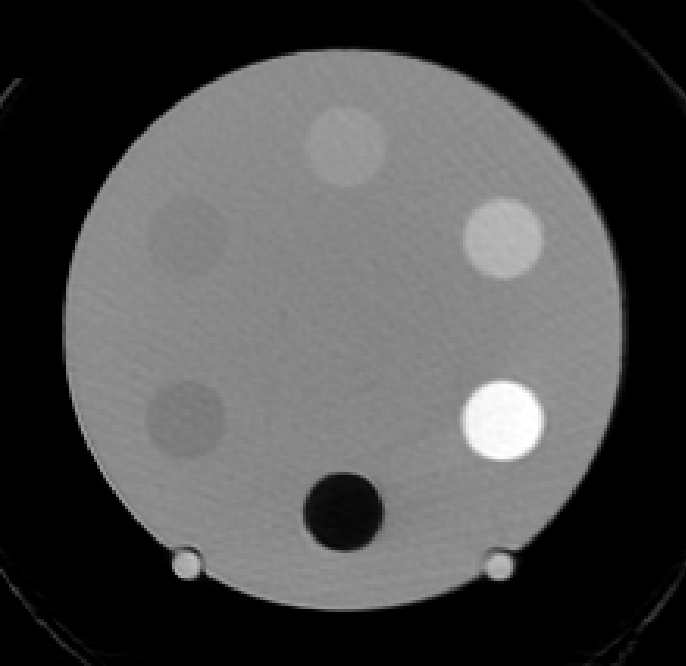}}
	\caption{Central slice of the CIRS phantom reconstruction including the CT linearity inserts used for determing the CNR and spatial resoultion (FWHM)}
	\label{fig:CIRS}
\end{figure}

\begin{figure}
	\centering
	\subfigure[SIRT 250 iterations]		{\includegraphics[width = 1.1in]{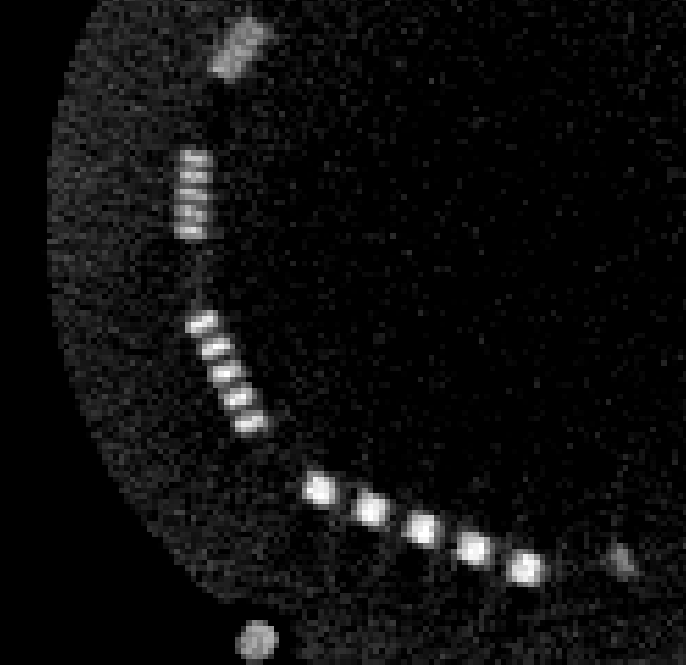}}
	\subfigure[lSIRT, highnoise model]	{\includegraphics[width = 1.1in]{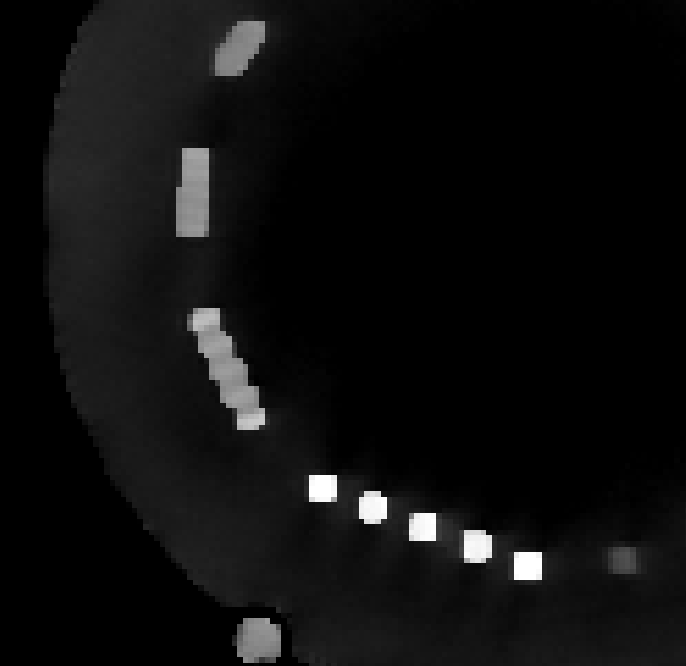}}
	\subfigure[lSIRT, patient model]	{\includegraphics[width = 1.1in]{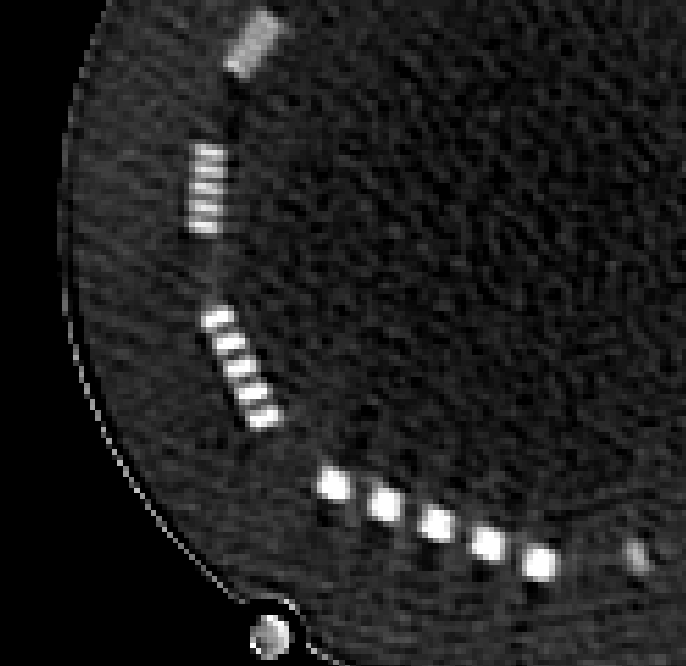}}
	\caption{Part of the Spatial Resolution Layer of the CIRS phantom.}
	\label{fig:CIRS_SRL}
\end{figure}

The results of the measured CIRS phantom are shown in Table~\ref{tab:CIRS}. Results of the reconstruction methods are displayed in Figure~\ref{fig:CIRS}. The models trained on the ellipse data show superior CNR compared to FBP, SIRT or patient trained methods. According to our metrics, the spatial resolution for all lSIRT methods is better than for FBP ($h = 0.8$ relative) and SIRT (250 iterations) but a visual inspection of the Spatial Resolution Layer of the phantom shows a better line pair differentiation for FBP and SIRT (Figure~\ref{fig:CIRS_SRL} a) compared to the ellips-trained lSIRT. This effect is clearly seen with the high-noise model (Figure~\ref{fig:CIRS_SRL} b). The patient trained lSIRT appears to be comparable with FBP and SIRT (Figure~\ref{fig:CIRS_SRL} c). 
It seems that the ellipse-trained lSIRT has not learned the shape of line-pairs and yields a inferior resolution for such details.

\begin{table}
	\caption{Reconstruction quality for measured CIRS scan}
	\centering
\begin{tabular}{l | c c}
Experiment & CNR & FWHM \\
           &   & (\SI{}{\centi\meter})\\
\hline
FBP (no filter)             &  1.81 & $\mathbf{0.14}$ \\
FBP (h=0.8)                 &  3.95 & 0.27 \\
SIRT (100 iterations)       & 21.76 & 0.40 \\
SIRT (250 iterations)       & 13.67 & 0.28 \\
SIRT (1000 iterations)      &  6.67 & 0.24 \\
lSIRT (low noise model)     & 27.16 & 0.25 \\
lSIRT (mid noise model)     & 33.17 & 0.24 \\
lSIRT (high noise model)    & $\mathbf{34.52 }$& 0.22 \\
lSIRT (Patient model)       & 17.11 & 0.23 \\
\hline
\end{tabular}\\
\label{tab:CIRS}
\end{table}

\section{Discussion}
From the results in Table~\ref{tab:triangles} and Table~\ref{tab:patient}, we can conclude that the lSIRT approach achieves better results than the classical reconstruction methods for 2D image reconstruction, which can be confirmed visually in Figures~\ref{fig:sl}, \ref{fig:patient} and \ref{fig:patienthn}. 
Furthermore, the lSIRT algorithms outperforms a U-net postprocessing approach and achieves competitive results compared to the LPD algorithm. The authors of the LPD algorithm have already shown \cite{Adler18} that U-net postprocessing is superior to total variation optimization allowing the same comparison for lSIRT.

The robustness of all our models was subsequently investigate for a class of out-of-distribution samples as depicted in Figure~\ref{fig:strange}, which suggests that the lSIRT algorithm is more robust to out-of-distribution samples when compared to other learned reconstruction methods. Such tests are an important part of testing before such models can be deployed into clinical practice.

One of the design goals of the lSIRT algorithm was to find a learned algorithm which is able to scale to clinically relevant problems and can be run with modest computational resources. In Table~\ref{tab:training} the memory requirements, which is well below the capabilities of commodity hardware, for each model and the number of parameters of the model are given. 

The results given in Table~\ref{tab:3D} illustrate that lSIRT can be scaled with success to 3D CBCT reconstruction. Figure~\ref{fig:cbct} provides several examples and clearly indicates that the model can be generalized beyond its training data.

When used to reconstruct real measured data of a physical phantom, the lSIRT algorithm provides good results, even though the number of projections and input resolution is very different from the data the model was trained with. Additionally, the forward projector used in this study, does not incorporate some of the physics (e.g.\ scatter) involved in the real measurement process. The quality of the reconstructions is given in Table~\ref{tab:CIRS}. The CNR is higher in the lSIRT algorithm, even though the line separation  (Figure~\ref{fig:CIRS_SRL}) appears to be slightly worse. It can be expected that the line separation improves when retraining the model with such relevant data. However, even in this case Figures~\ref{fig:strange}, \ref{fig:cbct} and \ref{fig:CIRS} illustrate that lSIRT can achieve good results for out-of-distribution samples. 

The advantages of the lSIRT approach are its ability to trade-off speed for memory which enables it to scale up to reconstruct CBCT images for clinically relevant problems. 
Furthermore, its robustness to changes in the projection operator (e.g., comparing simulations with real measurements) and its ability to reconstruct out-of-distribution samples is a clear advantage and implies that a single model can be applied to a wide range of scanning protocols and parts of the body. 

Currently, one of the disadvantages is the reconstruction time which is significantly longer than FBP, however, comparable to SIRT. Due to the construction of the algorithm, the reconstruction time can effectively be lowered by appropriate parallellization. Our implementation was not optimized for speed.

\section{Conclusion}
We have introduced the learned SIRT (lSIRT) algorithm, an algorithm for the reconstruction of CBCT scans. Our algorithm does not require a significant amount of memory during training and can be trained on CBCT problems of sizes of $128^3$ and $256^3$ voxels.
lSIRT takes a step to bring the enhanced image quality of deep-learning reconstruction techniques towards large 3D image reconstruction problems. Additionally, the method is shown to be flexible and relatively robust to the input data.


%






%
\bibliography{biblio}{}
\bibliographystyle{plain}
\end{document}